\documentclass[floats,floatfix,amssymb,prd,twocolumn,superscriptaddress,nofootinbib,preprintnumbers]{revtex4-1}

\usepackage{subcaption,ragged2e}
\usepackage{float}
\DeclareCaptionJustification{justified}{\justifying}
\usepackage{amssymb,amsmath,verbatim,mathtools,needspace,enumitem,etoolbox,graphicx,microtype,afterpage,bm}
\usepackage{tikz}
\usetikzlibrary{shapes.misc, positioning, arrows.meta}
\usepackage{framed}

\usepackage{hyperref}
\definecolor{linkcolor}{rgb}{0.0, 0.47, 0.75}
\definecolor{citecolor}{rgb}{1.0, 0.5, 0.0}
\hypersetup{
  linkcolor  = linkcolor,
  citecolor  = linkcolor,
  urlcolor   = linkcolor,
  colorlinks = true
}
\usepackage{multirow,pifont,lmodern,float,tabularx,booktabs}
\usepackage[all]{hypcap}
\usepackage[T1]{fontenc}
\usepackage[utf8]{inputenc}
\usepackage{cleveref}
\usepackage{algpseudocode}

\usepackage{float}

\newfloat{alg}{htbp}{loa}

\newenvironment{algorithm*}[1][htbp]{%
    \begin{figure*}[#1]%
    \addtocounter{figure}{-1}
    \refstepcounter{alg}%
    \captionsetup{
        name=ALG.,
        labelsep=colon
    }%
    \hrule height 0.8pt \kern\medskipamount 
}{%
\kern\medskipamount \hrule height 0.8pt 
\end{figure*}%
}

\renewcommand{\arraystretch}{1.4}

\allowdisplaybreaks
\newcommand{\bxobs}{{\mathbf x_\text{obs}}}

\newcommand{\bz}{\mathbf z}

\newcommand{\bx}{\mathbf x}

\usepackage{xcolor}
\begin{document}
\title{Dynamic SBI: Round-free Sequential Simulation-Based Inference\\ with Adaptive Datasets}

\author{Huifang Lyu}
\email{h.lyu@uva.nl}
\thanks{ORCID: \href{https://orcid.org/0000-0003-0008-3961}{0000-0003-0008-3961}}
\affiliation{GRAPPA Institute, Institute for Theoretical Physics Amsterdam,\\
University of Amsterdam, Science Park 904, 1098 XH Amsterdam, The Netherlands}

\author{James Alvey}
\email{jbga2@cam.ac.uk}
\thanks{ORCID: \href{https://orcid.org/0000-0003-2020-0803}{0000-0003-2020-0803}}
\affiliation{Kavli Institute for Cosmology Cambridge, Madingley Road, Cambridge CB3 0HA, United Kingdom}
\affiliation{Institute of Astronomy, University of Cambridge, Madingley Road, Cambridge CB3 0HA, United Kingdom}

\author{Noemi Anau Montel}
\email{noemiam@mpa-garching.mpg.de}
\thanks{ORCID: \href{https://orcid.org/0000-0001-8985-7534}{0000-0001-8985-7534}}
\affiliation{Max-Planck-Institut für Astrophysik, Karl-Schwarzschild-Str.\ 1, 85748 Garching, Germany}

\author{Mauro Pieroni}
\email{mauro.pieroni@csic.es}
\thanks{ORCID: \href{https://orcid.org/0000-0003-0665-266X}{0000-0003-0665-266X}}
\affiliation{Instituto de Estructura de la Materia (IEM), CSIC, Serrano 121, 28006 Madrid, Spain}

\author{Christoph Weniger}
\email{c.weniger@uva.nl}
\thanks{ORCID: \href{https://orcid.org/0000-0001-7579-8684}{0000-0001-7579-8684}}
\affiliation{GRAPPA Institute, Institute for Theoretical Physics Amsterdam,\\
University of Amsterdam, Science Park 904, 1098 XH Amsterdam, The Netherlands}

\begin{abstract}
\noindent 
Simulation-based inference (SBI) is emerging as a new statistical paradigm for addressing complex scientific inference problems. By leveraging the representational power of deep neural networks, SBI can extract the most informative simulation features for the parameters of interest. Sequential SBI methods extend this approach by iteratively steering the simulation process towards the most relevant regions of parameter space. This is typically implemented through an algorithmic structure, in which simulation and network training alternate over multiple rounds. This strategy is particularly well suited for high-precision inference in high-dimensional settings, which are commonplace in physics applications with growing data volumes and increasing model fidelity. Here, we introduce dynamic SBI, which implements the core ideas of sequential methods in a round-free, asynchronous, and highly parallelisable manner. At its core is an adaptive dataset that is iteratively transformed during inference to resemble the target observation. Simulation and training proceed in parallel: trained networks are used both to filter out simulations incompatible with the data and to propose new, more promising ones. Compared to round-based sequential methods, this asynchronous structure can significantly reduce simulation costs and training overhead. We demonstrate that dynamic SBI achieves significant improvements in simulation and training efficiency while maintaining inference performance. We further validate our framework on two challenging astrophysical inference tasks: characterising the stochastic gravitational wave background and analysing strong gravitational lensing systems. Overall, this work presents a flexible and efficient new paradigm for sequential SBI.
\end{abstract}

\maketitle

\preprint{}

\section{Introduction}\label{sec:intro}

\begin{figure*}
    \centering
    \includegraphics[width=\linewidth]{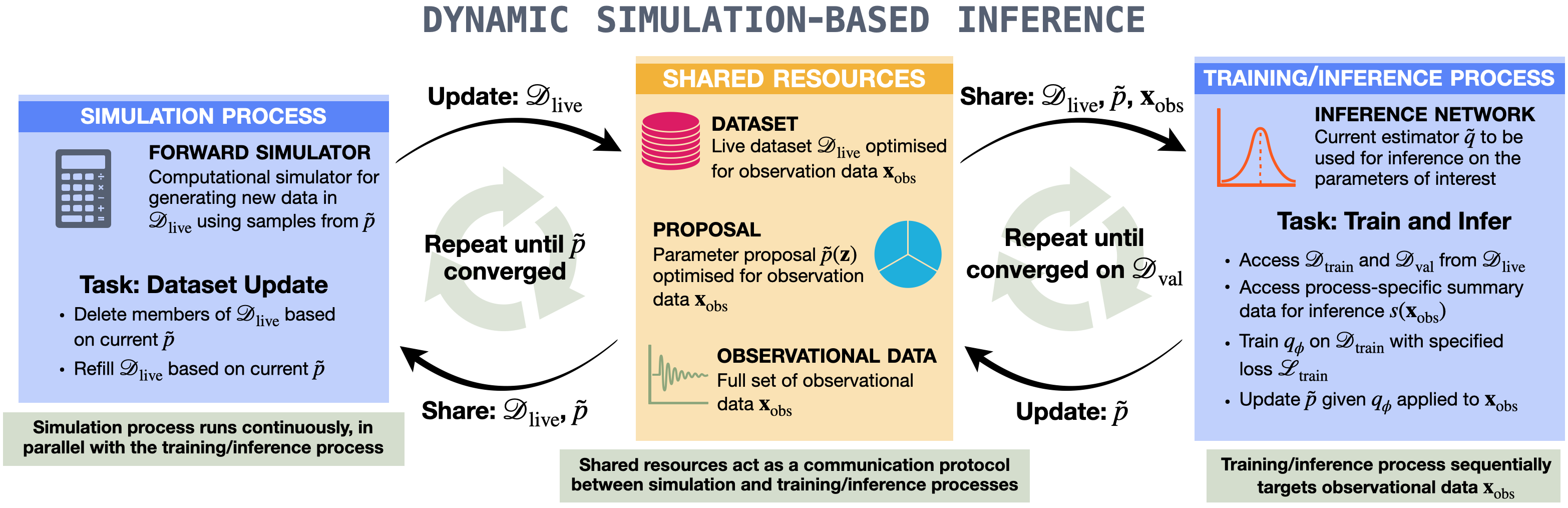}
    \caption{\textbf{Dynamic simulation-based inference:} Summary figure detailing the dynamic SBI framework. The framework splits into two core process types: a) a \emph{simulation} process which updates (via deletion/re-simulation) a live dataset $\mathcal{D}_{\mathrm{live}}$ to reflect the current state of the proposal $\tilde{p}(\bz)$; and b) a \emph{training/inference} process where a posterior estimate $q_\phi$ is optimised on the current state of the dataset $\mathcal{D}_{\mathrm{live}}$, and the proposal $\tilde{p}$ is subsequently updated to sequentially target some observed data $\bxobs$. These two processes run in parallel, and communicate via a set of \emph{shared resources}, which includes the live dataset $\mathcal{D}_{\mathrm{live}}$, the parameter proposal $\tilde{p}(\bz)$, and the observational data $\bxobs$.}
    \label{fig:summary}
\end{figure*}

\noindent 
Statistical inference plays a central role in modern science, ultimately allowing us to connect physical models and their parameters to observational data. However, in many scenarios, this inference step can pose a significant computational and methodological bottleneck. Reasons include, \textit{e.g.}, slow convergence of stochastic sampling techniques, practical but not fully accurate approximations to the data likelihoods, or ambiguous parametric models for complex scenarios or systems.  In recent years, simulation-based inference (SBI) (see Ref.~\cite{Cranmer2020frontier} for an early review) has emerged as a promising set of methods to potentially overcome these challenges in settings where forward simulators are available. 

Broadly speaking, SBI techniques can be categorised into a number of algorithm families such as: Neural Posterior Estimation (NPE)~\cite{Papamakarios2016npe,Zeghal2022differentiable}, which directly learns the posterior distribution; Neural Likelihood Estimation (NLE)~\cite{Papamakarios2019snl,Alsing:2019xrx}, which learns a neural representation of the likelihood; and Neural Ratio Estimation (NRE)~\cite{Brehmer:2018hga,Stoye:2018ovl,Hermans2020ratio,Durkan2020contrastive,Rozet2021amnre,Delaunoy2022balanced}, which constructs an estimator for the likelihood-to-evidence ratio. Additional examples include Ref.~\cite{Gloeckler2024allinone}, which trains a transformer model on the joint distribution of parameters and data, and Ref.~\cite{Wildberger2023flowmatching}, which learns the posterior with flow matching. These approaches can be further categorised as either \emph{amortised} (see Ref.~\cite{ZammitMangion2025neural} for a review), where a neural network is trained over the entire prior space and then applicable to multiple observations, or \emph{sequential}, where instead simulations are iteratively guided to regions of interest and networks retrained to maximise their efficacy for a specific, target observation. This can be particularly important for scenarios where amortisation is difficult or impossible, for example, because the target posterior is too high-dimensional, too high-precision, or generating sufficiently many simulations is too costly (termed an ``amortisation gap'' in Ref.~\cite{ZammitMangion2025neural}).

Focusing on this latter class of iterative methods, modern implementations of the core SBI algorithms all have sequential analogues, see Refs.~\cite{Greenberg2019apt,Lueckmann2017flexible,Deistler2022truncated,Greisemer2024active} (sequential NPE methods),~\cite{Papamakarios2019snl,Alsing:2019xrx} (sequential NLE algorithms), and~\cite{Miller:2021hys,Miller:2022shs} (sequential NRE setups). Naturally, each method differs somewhat in the specific benefits/trade-offs it offers. For example, the mechanisms by which the different implementations refine the proposal distributions for new simulations range from Bayesian optimisation and acquisition~\cite{Alsing:2019xrx}, to proposal invariance~\cite{Greenberg2019apt}, and prior truncation~\cite{Deistler2022truncated,Miller:2021hys}. Nonetheless, they share the common feature that they can be a highly effective choice for computationally expensive simulators. This has been demonstrated widely across concrete applications in astrophysics, cosmology, engineering, ecology, and beyond (see Refs.~\cite{Alsing2023optimal,Hull2024watershed,AnauMontel:2022ppb,Crisostomi:2023tle,Makinen:2021nly,LSSTDarkEnergyScience:2024gzn,Dupourque:2025vqe,Nguyen:2024yth,Pacilio:2024qcq,Khullar2022digs,Savchenko:2025jzs,Cole:2025sqo,Konar:2024mrz,Bhardwaj:2023xph,Alvey:2023naa,Alvey:2023npw,Eckner:2025waa,Karchev:2022xyn,Mediato-Diaz:2025oqv,Marlier2024grasping,Vilchez:2024qnw,vonWietersheim-Kramsta:2024cks,Wagner-Carena:2024axc,Coogan:2022cky,Karchev:2022xyn,Cole:2021gwr} for a broad selection of sequential SBI applications). Another important feature that this class of algorithms typically (although not exclusively, see, \emph{e.g.},~\cite{Alsing:2019xrx, Wagner-Carena:2024axc}) share is structuring the iterative update of proposals and simulation generation into discrete \emph{rounds} during which estimator networks are re-trained on progressively focused regions of the posterior distribution.

Round-based sequential inference methods may exhibit computational inefficiencies.  These fall into two categories: a) the redundant use/generation of simulation data, and b) unnecessary training steps, both of which contribute to a theoretical excess in inference cost. This ultimately stems from an inherent trade-off in round-based methods between simulation generation and training time. Specifically, within a ``successful'' round (one where sufficient information is gathered to meaningfully update the parameter proposal distribution), by the time the inference networks converge, many simulations become highly improbable given the observation under consideration. This suggests the possibility of introducing more efficient sequential inference algorithms that utilise information as it becomes available, rather than waiting for round completion.

The general goal of the present work is to bridge this gap by proposing a framework of \emph{dynamic} sequential SBI. Two key innovations position our proposed algorithm structure as a hybrid between amortised methods that harness active learning~\cite{Settles2009active} techniques to choose new parameter points for simulations, and round-based sequential techniques that update proposal distributions and re-train networks in discrete steps. The first core property is viewing the training dataset as a continuous resource that can be updated while training. This idea is linked to the second main development, which proposes removing disfavoured parameter points in the training dataset as soon as the inference network provides sufficient ``evidence'' (broadly defined, see the specific algorithms below for details), and adding new samples based on the current training state. These innovations open up a number of important benefits, including: the ability to asynchronously train proposal distributions for generating new simulations and removing the necessity to train multiple networks from scratch to sequentially target one observation. To summarise, the key gap we identify and address is the lack of a truly ``online'' framework where the distinction between data generation and model training is removed, allowing for continuous, dynamic adaptation of both the dataset and the proposal distributions.

Our work builds upon an extensive literature on adaptive and sequential methods in the field of Bayesian computation. For example, the core idea of refining proposals or datasets in light of new information is central to the field of active learning~\cite{Settles2009active}, which has found successful application in SBI (see, \emph{e.g.}, Ref.~\cite{Alsing:2018eau} for an example in the context of cosmology). There are also significant conceptual similarities with nested sampling~\cite{Skilling2006nested,Handley:2015fda,Ashton:2022grj}. In particular, nested sampling proceeds by sequentially compressing a set of ``live'' points (the training dataset in this paper) to lie within a monotonically increasing iso-likelihood contour by discarding the lowest likelihood point (at least in the vanilla implementation~\cite{Skilling2006nested}) in the set and resampling a new point with a higher likelihood. In our context, we turn this procedure into discarding and updating all simulation points that do not contain enough information, given the current inference state. This feature naturally couples the rate of prior volume compression to the amount of information obtained by the trained networks. Finally, the sequence of iterative proposal and posterior distributions generated during our training process can be naturally interpreted through the lens of adaptive annealing or bridging distributions, which are foundational concepts in Sequential Monte Carlo (SMC) methods~\cite{Buchholz2018smc}.

The rest of the paper is structured as follows: in Section~\ref{sec:online}, we present the high-level dynamic SBI algorithm structure, and discuss specific implementations that can be used in practice. Then, in Section~\ref{sec:benchmarking}, we demonstrate the key performance benefits surrounding simulation and training costs that we claimed above across a challenging benchmark scenario. In Section~\ref{sec:physics}, we apply the set of algorithms to two physics cases in the field of astrophysics (stochastic gravitational wave background recovery and the analysis of strong lensing images). Finally, we conclude in Section~\ref{sec:conclusions}.

\section{Dynamic Simulation-based Inference}\label{sec:online}

\noindent In this section, we lay out the core framework, mathematical formulation, and specific algorithmic implementations that underpin our dynamic SBI (DS) methodology. This is illustrated schematically in Fig.~\ref{fig:summary}, and detailed at the general algorithm level in Algorithm~\ref{alg:NPE}.

\begin{algorithm*}
\caption{\textbf{Algorithm} -- Dynamic Simulation-Based Inference (DS)}
\begin{algorithmic}[1]
\State \textbf{Input:}
\State \hspace{\algorithmicindent} (Implicit) prior distribution $p(\bz)$
\State \hspace{\algorithmicindent} (Implicit) likelihood model/simulator $p(\bx \mid \bz)$
\State \hspace{\algorithmicindent} Neural density estimator $q_\phi(\bz \mid \bx)$
\State \hspace{\algorithmicindent} Number of simulations $N$ in $\mathcal{D}_\mathrm{live}$
\State \hspace{\algorithmicindent} Observation $\bxobs$
\State \hspace{\algorithmicindent} Proposal update function $\tilde{p}(\bz) \leftarrow \texttt{proposal\_update}[\tilde{p}(\bz), \bxobs, q_\phi(\bz \mid \bx)]$
\State \hspace{\algorithmicindent} Dataset update function $\mathcal{D}_{\mathrm{live}} \leftarrow \texttt{data\_update}[\mathcal{D}_{\mathrm{live}}, \bxobs, \tilde{p}(\bz)]$
\State \hspace{\algorithmicindent} Convergence criterion $\texttt{converged}[\mathcal{D}_{\mathrm{live}}, \bxobs, q_\phi(\bz \mid \bx), \tilde{p}(\bz)]$
\State \textbf{Output:} Trained neural network $q_\phi(\bz \mid \bx)$
\State
\State \textbf{Initialisation:}
\State Initialise proposal distribution $\tilde{p}(\bz) = p(\bz)$
\State Initialise training dataset $\mathcal{D}_{\mathrm{live}} \gets \emptyset$
\For{$i = 1$ to $N$}
    \State Sample $\bz^{(i)}$ from the prior $p(\bz)$
    \State Generate synthetic data $\bx^{(i)}$ from the model $p(\bx \mid \bz)$
    \State Add the data $(\bx^{(i)}, \bz^{(i)})$ to the dataset $\mathcal{D}_{\mathrm{live}}$
\EndFor
\State Initialise neural network $q_\phi(\bz \mid \bx)$ with parameters $\phi$
\State
\State \textbf{Main Algorithm:}
\While{not $\texttt{converged}[\mathcal{D}_{\mathrm{live}}, \bxobs, q_\phi(\bz \mid \bx), \tilde{p}(\bz)]$} \textbf{(in parallel)}
    \State \textbf{[Parallel Process 1] Training Loop:}
    \For{$i = 1$ to $N_\mathrm{batches}$}
        \State Sample mini-batch $\mathcal{B}_i$ from $\mathcal{D}_{\mathrm{live}}$
        \State Compute the training loss: $\mathcal{L}_\mathrm{train}[\phi, \mathcal{B}_i]$
        \State Update network parameters $\phi$ using gradient descent on $\mathcal{L}_\mathrm{train}[\phi, \mathcal{B}_i]$
    \EndFor
    \State Update proposal distribution $\tilde{p}(\bz) \leftarrow \texttt{proposal\_update}[\tilde{p}(\bz), \bxobs, q_\phi(\bz \mid \bx)]$
    \State
    \State \textbf{[Parallel Process 2] Update Step:}
    \State Update dataset $\mathcal{D}_{\mathrm{live}} \leftarrow \texttt{data\_update}[\mathcal{D}_{\mathrm{live}}, \bxobs, \tilde{p}(\bz)]$
\EndWhile
\State
\State \textbf{Return} Trained neural network $q_\phi(\bz \mid \bx)$
\end{algorithmic} \label{alg:NPE}
\end{algorithm*}

\subsection{Parallel Asynchronous Inference Framework}

\noindent 
At the most basic level, there are two key differences between our algorithm (Algorithm~\ref{alg:NPE}) and typical sequential SBI approaches~\cite{Papamakarios2019snl,Lueckmann2017flexible,Greenberg2019apt}. First, we treat the training (and validation) datasets as dynamic resources. In other words, unlike in round-based approaches, where the training dataset remains fixed during each round, we have a prescription for a dataset update (\texttt{data\_update} in Algorithm~\ref{alg:NPE}) that can be run continuously during training. Secondly, the proposal is continuously updated (again, as described in the meta algorithm via the \texttt{proposal\_update} function). The exact specification for the proposal, and how it relates to current estimates of the posterior, is algorithm-dependent, but, in general, it allows for the dataset to be immediately refilled with newly proposed points that sequentially target the observational data under consideration. Overall, this procedure aims to achieve two goals: i) a dataset point is rejected as soon as it becomes uninformative during training, and ii) new informed data is generated and used to update the network as soon as possible. Together, these guidelines ensure that simulations are generated ``on demand'' and as little time as possible is spent training networks on uninformative data. 

We now introduce some specific definitions that we will use throughout this work. Independent of the algorithm, we consider inference problems where we aim to approximate some true posterior $p(\bz \mid \bxobs)$ for some model parameters $\bz$ given some observed data $\bxobs$. We assume we can draw prior samples $\bz \sim p(\bz)$ and generate simulated data $\bx$ from a (simulated) data likelihood $p(\bx \mid \bz)$. In the initialisation phase of each algorithm, we populate our (live, dynamical) dataset $\mathcal{D}_{\mathrm{live}}$ with samples from the joint distribution $p(\bx, \bz) = p(\bx \mid \bz) p(\bz)$, and initialise our (again, dynamical) proposal $\tilde{p}(\bz)$ to the prior distribution $p(\bz)$. In any implementation, the goal is to train a neural network estimator $q_\phi(\bz \mid \bx)$ that accurately approximates the true posterior $p(\bz \mid \bxobs)$ (or a suitable proxy) for the observation of interest $\bxobs$. This is achieved by training $q_\phi$ using gradient descent methods on some algorithm-dependent loss $\mathcal{L}_\mathrm{train}$, iterating through mini-batches of the current state of the live dataset $\mathcal{D}_{\mathrm{live}}$. This training is continued until some convergence criterion (the boolean flag \texttt{converged}[$\mathcal{D}_{\mathrm{live}}, \bxobs, q_\phi(\bz \mid \bx), \tilde{p}(\bz)$] in Algorithm~\ref{alg:NPE}) is met. This criterion can again depend on the algorithm, and might be a series of different conditions that must all be met, such as \emph{e.g.}, the dataset must remain unchanged, the loss on some validation set is stable, or other related quality control measures.

\subsection{Modified Proposals and Neural Posterior Estimation}

\noindent In the remainder of this section, we will present two specific versions of the meta algorithm (Algorithm~\ref{alg:NPE}) that we propose in this work. These follow very closely (and are named accordingly) the taxonomy of sequential NPE algorithms already present in the literature: SNPE-A~\cite{Papamakarios2016npe}, SNPE-B~\cite{Lueckmann2017flexible}, and SNPE-C (or APT)~\cite{Greenberg2019apt}, adapted for the dynamic setting. Full algorithm blocks for each algorithm are provided in the appendix. Before discussing these, however, we will first briefly review the core NPE algorithm, specifically the impact of proposal distributions. 

The training objective in our dynamic framework (specified by $\mathcal{L}_\mathrm{train}$), whilst somewhat algorithm dependent, relying on the particular proposal settings, broadly follows the formulation of neural posterior estimation (NPE)~\cite{Papamakarios2016npe,Greenberg2019apt,Lueckmann2017flexible}. In NPE, the target loss function is given by the negative log-likelihood of the posterior estimator $q_\phi(\bz \mid \bx)$, averaged over the joint distribution, \emph{i.e.},
\begin{align}\label{eq:NPE_loss}
    \mathcal{L}_\mathrm{NPE}(\phi) &= -\mathbb{E}_{p(\bz, \bx)}[\log q_\phi(\bz \mid \bx)] \nonumber \\ 
    &= - \int{\mathrm{d}\bz \, \mathrm{d}\bx \, p(\bz, \bx) \log q_\phi(\bz \mid \bx)} \;.
\end{align}
In practice, this is always approximated via a Monte Carlo average:
\begin{equation} \label{eq:NPE_loss_sum}
    \mathcal{L}_\mathrm{NPE}(\phi) \simeq -\cfrac{1}{N}\sum_{i = 1}^{N}{\log q_\phi(\bz_i \mid \bx_i)} \;,
\end{equation}
where training data is generated from a forward model, $\bx_i, \bz_i \sim p(\bx \mid \bz) p(\bz)$ for $i=1, \dots, N$. Optimising the loss in Eq.~\eqref{eq:NPE_loss} trains the network $q_\phi$ to directly approximate the true posterior $q_\phi(\bz \mid \bx) \simeq p(\bz \mid \bx)$. One way to see this is the fact that minimising Eq.~\eqref{eq:NPE_loss} w.r.t. $\phi$ is equivalent (up to an additive constant independent of $q_\phi$) to minimising the Kullback-Leibler (KL) divergence between the true posterior $p(\bz \mid \bx)$ and the learned density estimator $q_\phi(\bz \mid \bx)$, averaged over draws from the marginal distribution of the data $p(\bx)$. 

In the context of sequential SBI, one iteratively replaces the prior $p(\bz)$ with a suitably chosen proposal distribution $\tilde p(\bz)$ that focuses simulation data on the observation of interest. If the standard NPE loss function in Eq.~\eqref{eq:NPE_loss} is used without modification, then the network learns to approximate a proposal-biased posterior:
\begin{equation}
    q_\phi(\bz \mid \bx) \simeq  \cfrac{1}{Z} p(\bx \mid \bz) \tilde p(\bz) \;,
\end{equation}
where $Z$ is a normalisation constant. Regardless of the proposal, the goal of all sequential SBI algorithms is still to recover the true posterior $p(\bz \mid \bx) \propto p(\bx \mid \bz) p(\bz)$. As such, each of the algorithm families (SNPE-A/B/C) has different methods to undo the effects of proposals that are not the prior.

For example, in the SNPE-A~\cite{Papamakarios2016npe} version of sequential SBI, the analysis is carried out in multiple rounds $r$, with proposal distributions initially set to be the prior
$\tilde p^{(r=1)}(\bz) = p(\bz)$ and then iterative approximations to the posterior evaluated on the observation of interest $\bxobs$
$\tilde p^{(r)}(\bz) = q_\phi^{(r-1)}(\bz \mid \bxobs)$.
After $r$ rounds of training on the loss specified in Eq.~\eqref{eq:NPE_loss}, the true posterior is related to the learned approximate posterior $q_\phi^{(r)}(\bz \mid \bx)$ via the following
\begin{equation}\label{eq:npe_a_result}
    p(\bz \mid \bx) \propto q_\phi^{(r)}(\bz \mid \bx) \cfrac{p(\bz)}{\tilde p^{(r)}(\bz)} \;.
\end{equation}
We see that this strategy typically learns a distribution $q_\phi^{(r)}$ that is narrower than the true posterior. This can be undone by taking samples from $q_\phi^{(r)}$ and reweighting them according to the ratio in the last term of Eq.~\eqref{eq:npe_a_result}.

Other variants of sequential SBI algorithms (specifically SNPE-B~\cite{Lueckmann2017flexible} and SNPE-C~\cite{Greenberg2019apt}) solve this problem of proposals differently. In particular, both of these algorithms modify the training loss $\mathcal{L}_\mathrm{train}$ itself: SNPE-B via importance weighting, and SNPE-C with a classification-inspired loss on atomic proposals. In both cases, the learned distribution $q_\phi$ \emph{directly} approximates the true posterior $q_\phi(\bz \mid \bx) \simeq p(\bz \mid \bx)$, requiring no reweighting during sampling. With this range of options in mind, we discuss below how we account for this proposal dependence of standard NPE methods in the dynamic SBI setting. 

\subsection{Algorithm: DS-A}
\label{sec:DS-A_algorithm}
\vspace{6pt}

\noindent The first dynamic SBI algorithm (called DS-A here) that we present is built on top of the SNPE-A~\cite{Papamakarios2016npe} framework. In the context of training, the algorithm learns an approximation $q_\phi$ to the ``proposal-biased'' posterior, $q_\phi(\bz \mid \bx) \sim \tilde{p}(\bz \mid \bx)$ with:
\begin{equation}
\tilde{p}(\bz \mid \bx) \propto \cfrac{\tilde{p}(\bz)}{p(\bz)}  p(\bz \mid \bx) \; .
\end{equation}
Here, $\tilde{p}(\bz)$ is the proposal distribution of parameters $\bz$ in the dataset $\mathcal{D}_{\mathrm{live}}$, $p(\bz)$ is the original prior distribution, and $p(\bz \mid \bx)$ is the true posterior. 

This is achieved by choosing the loss function $\mathcal{L}_\mathrm{train}$ that matches the standard NPE loss given above, \emph{i.e.},
\begin{equation}
    \displaystyle \mathcal{L}_\mathrm{train} \equiv \mathcal{L}_{\mathrm{DS-A}}(\phi)= -\cfrac{1}{N}\sum_{i=1}^{N}\log q_{\phi}(\bz_i\mid \bx_i)\; ,
\end{equation}
with $\bx_i, \bz_i \sim p(\bx \mid \bz) \tilde{p}(\bz)$. With this formulation, it remains to make a choice for the proposal distribution $\tilde{p}(\bz)$.

The \emph{traditional} SNPE-A~\cite{Papamakarios2016npe, Greenberg2019apt} algorithm typically assumes a \textit{Gaussian} proposal. This choice enables a closed-form test-time correction
$q_{\phi}(\bz\mid \bxobs)\,p(\bz)/\tilde p(\bz)$ to be analytically normalizable and directly samplable for every round, but it also limits flexibility: Gaussian proposals struggle with multi-modality or heavy tails and can induce mode collapse or poor coverage when the proposal has weak overlap with the true posterior.

Given these limitations, a tempting alternative is to use the current state of the trained flow as the proposal distribution, \emph{i.e.}, $\tilde p(\bz) = q_\phi(\bz \mid \bxobs)$. However, this ansatz would generate a feedback loop that would formally shrink the learned distribution to a point, introducing significant issues when trying to re-weight samples.

To keep these effects under control, we instead choose to have a proposal distribution that is proportional to a tempered version of the likelihood, determined by a factor $\gamma$.
\begin{equation}
    \tilde p(\bz) \propto p(\bxobs \mid \bz)^\gamma p(\bz)\;.
\end{equation}
This is achieved by the ansatz 
\begin{equation}
    \tilde p(\bz) = 
    \left(\cfrac{q_\phi(\bz \mid \bxobs)}{p(\bz)}
    \right)^{\frac{\gamma}{1+\gamma}}
    p(\bz)\;,
\end{equation}
which can be confirmed by plugging in $q_\phi(\bz \mid \bx) \propto p(\bx \mid \bz) \tilde{p}(\bz)$ from above. We note that in the limit $\gamma \to \infty$, this choice of proposal reduces to the situation discussed above, $\tilde p(\bz) \to q_\phi(\bz \mid \bxobs)$, in which case the equilibrium proposal distribution becomes very peaked.

To obtain the true posterior $p(\bz\mid \bx)$ from the learned model $q_\phi(\bz \mid \bxobs)$, we apply an importance correction,
\begin{equation}
    p(\bz \mid \bxobs) \propto \frac {p(\bz)}{p'(\bz)}
    q_\phi(\bz \mid \bxobs) \;,
\end{equation}
where $p'(\bz)$ denotes the \emph{actual} distribution of training data towards the end of the network training. Note that although $p'(\bz)$ approximates the target $\tilde p(\bz)$, those distributions generally differ, because the training data state is also affected by previous states of the proposal $\tilde p(\bz)$.  One can deal with this in two ways.  First, one can train another network to approximate the current training data state, $q_\psi(\bz) \approx p'(\bz)$, and use this for the above correction. Alternatively, assuming that $p'(\bz) \simeq \tilde p(\bz)$, one can obtain the closed-form solution.
\begin{equation}
p(\bz\mid \bx)\;\propto\;\cfrac{p(\bz)}{\tilde p(\bz)}\,q_{\phi}(\bz\mid \bx) \propto \left(\cfrac{p(\bz)}{q_\phi(\bz \mid \bxobs)}\right)^{\frac{\gamma}{1 + \gamma}} q_\phi(\bz \mid \bx)\; .
\end{equation}
We follow here the latter approach. At the sample level, we can obtain $\bz \sim p(\bz \mid \bxobs)$ by first sampling $\bz \sim q_\phi(\bz \mid \bxobs)$ and then reweighting by the ratio $p(\bz)/\tilde p(\bz)$~\cite{Papamakarios2016npe}. 

The final component of DS-A concerns updating the training dataset. We remove a point from $\mathcal{D}_\mathrm{live}$ whenever
\begin{equation}\label{eq:rthreshold}
   r \;\equiv\; \ln \frac{q_\phi(\bz \mid \bxobs)}{q'_{\phi}(\bz \mid \bxobs)} \;<\; r_{th}\,,
\end{equation}
where $q'_{\phi}(\bz \mid \bxobs)$ comes from the $q_\phi(\bz \mid \bxobs)$ at the time the point was generated.  This criterion is convenient to evaluate because $q'_{\phi}(\bz \mid \bxobs)$ is stored upon creation and $q_\phi(\bz\mid\bxobs)$ can be computed efficiently throughout training. Conceptually, it acts as loose proxy for the log-ratio of the current proposal distribution and the proposal distribution at sampling time. In the experiments below, we set the threshold to $r_{th} = -20$, an empirically robust choice corresponding to $q_\phi(\bz\mid\bxobs)/q'_{\phi}(\bz \mid \bxobs)\lesssim 2\times10^{-9}$, thereby pruning points that carry negligible information under the current posterior.

\vspace{6pt}

\noindent To summarise and connect to Algorithm~\ref{alg:NPE}, the DS-A version of dynamic SBI implements the following method choices.
\begin{itemize}
    \item \texttt{loss}:
    Train $q_{\phi}(\bz\mid\bx)$ by minimizing the unweighted NPE objective on $(\bx,\bz)\sim p(\bx\mid\bz)\,\tilde p(\bz)$,
    $\displaystyle \mathcal{L}_{\mathrm{DS-A}}(\phi)= -\cfrac{1}{N}\sum_{i=1}^{N}\log q_{\phi}(\bz_i\mid \bx_i)$.
    At inference, recover the posterior by importance correction,
    $\displaystyle p(\bz\mid\bx)\propto \cfrac{p(\bz)}{\tilde p(\bz)}\,q_{\phi}(\bz\mid\bx)$.
    
    \item \texttt{proposal\_update}:
    Use a tempered proposal
    $\displaystyle \tilde p(\bz)\propto\Big(\cfrac{q_{\phi}(\bz\mid\bx_{\mathrm{obs}})}{p(\bz)}\Big)^{\frac{\gamma}{1+\gamma}}p(\bz)$ with a tempering factor $\gamma$.
    This avoids the over-concentration caused by the naive choice $\tilde p(\bz)=q_{\phi}(\bz\mid\bx_{\mathrm{obs}})$.
    
    \item \texttt{data\_update}:
    Simulate fresh pairs $(\bx,\bz)\sim p(\bx\mid\bz)\,\tilde p(\bz)$ to refresh $\mathcal{D}_{\mathrm{live}}$.
    Prune legacy samples with $r=\ln\!\big[q_{\phi}(\bz\mid\bx_{\mathrm{obs}})/q'_{\phi}(\bz \mid \bxobs)\big]<r_{th}$.
    
    \item \texttt{converged}:
    Convergence is declared via an early–stopping criterion: if the validation loss fails to attain a new minimum for a specified number of consecutive epochs. The network is then reset to the state with the best validation loss.
\end{itemize}
In the below experiments, we will adopt the hyper parameter choices $\gamma = 0.5$ and $r_{th} = -20$. The implementation details of DS-A within the meta-algorithm are described fully in Algorithm~\ref{alg:OS-A}, and the performance of this algorithm is explored in Section~\ref{sec:benchmarking}. 

\subsection{Algorithm: DS-B}
 
\noindent In the dynamic SBI variant of SNPE-B~\cite{Lueckmann2017flexible} (called DS-B here), the fact that the distribution describing the parameters $\bz$ in the dataset $\mathcal{D}_{\mathrm{live}}$ may not be the prior $p(\bz)$ is accounted for using importance weights. In particular, as pointed out in Refs.~\cite{Papamakarios2016npe,Lueckmann2017flexible}, the expectation in Eq.~\eqref{eq:NPE_loss_sum} can be re-written as: 
\begin{equation}
    \mathcal{L}_\mathrm{NPE}(\phi) = -\cfrac{1}{N}\sum_{i=1}^{N}\left[\cfrac{p(\bz_{i})}{p'(\bz_{i})} \log q_\phi(\bz_{i} \mid \bx_{i})\right] \; ,
\end{equation}
where $p'(\bz)$ is the distribution of parameters $\bz$ in the dataset $\mathcal{D}_{\mathrm{live}}$.\footnote{Note here that, technically, this \emph{is not} necessarily the prior $p(\bz)$, the proposal $\tilde{p}(\bz)$, or the posterior $q_\phi(\bz \mid \bx)$, but a hybrid distribution that reflects the current state of the dataset. This may have had parameters deleted, but may also still contain prior samples.} In other words, the importance weights $w(\bz) = p(\bz) / p'(\bz)$ are applied to each parameter point $\bz$ such that the trained density estimator $q_\phi(\bz \mid \bx)$ still converges directly to the posterior $p(\bz \mid \bx)$. 

In the dynamic implementation, detailed in Algorithm~\ref{alg:OS-B}, we implement the specifics of the meta-algorithm in the following ways. For the proposal update, we simply take the current posterior estimator evaluated on the observed data $\bxobs$ as the new proposal $\tilde{p}(\bz) \leftarrow q_\phi(\bz \mid \bxobs)$. Then, to determine the points to delete in the current live dataset $\mathcal{D}_{\mathrm{live}}$, we carry out the following procedure. Firstly, we define a quantile level $\alpha_B$ for the values of the log-proposal density $\log \tilde{p}(\bz)$ and compute the corresponding threshold value $c_B$.\footnote{In practice this is achieved by sampling directly $\bz_i \sim q_\phi(\bz \mid \bxobs)$, computing $l_i = \log q_\phi(\bz_i \mid \bxobs)$ and then sorting the set of values $\{l_i\}_{i = 1}^{K}$ for suitably large value of $K$ that is sufficient to estimate the $\alpha_B$ contour.} We then delete any points $\bz_i$ in $\mathcal{D}_{\mathrm{live}}$ with $\log \tilde{p}(\bz_i) < c_B$ and re-simulate a new point $\bz_\mathrm{new} \sim \tilde{p}(\bz)$. Since this point is not drawn from the prior, we also store the importance weight $w_\mathrm{new} = p(\bz_\mathrm{new})/\tilde{p}(\bz_\mathrm{new})$, which measures the relative probability of how likely the point was to be generated from the proposal compared to a prior sample. This weight remains fixed for the rest of the time that the parameter point is maintained in $\mathcal{D}_{\mathrm{live}}$. Finally, we implement the importance weighted training loss given by:
\begin{equation}
    \displaystyle \mathcal{L}_\mathrm{train} \equiv \mathcal{L}_\mathrm{DS-B}(\phi) = -\cfrac{1}{N}\sum_{i = 1}^N{w_i \log q_\phi(\bz_i \mid \bx_i)} \; ,
\end{equation}
where $w_i$ is computed as above for each parameter set $\bz_i$ and is initialised to $w_i = 1$ for the initial set of prior samples.

\begin{figure*}[t]
    \centering
    \includegraphics[width=\linewidth]{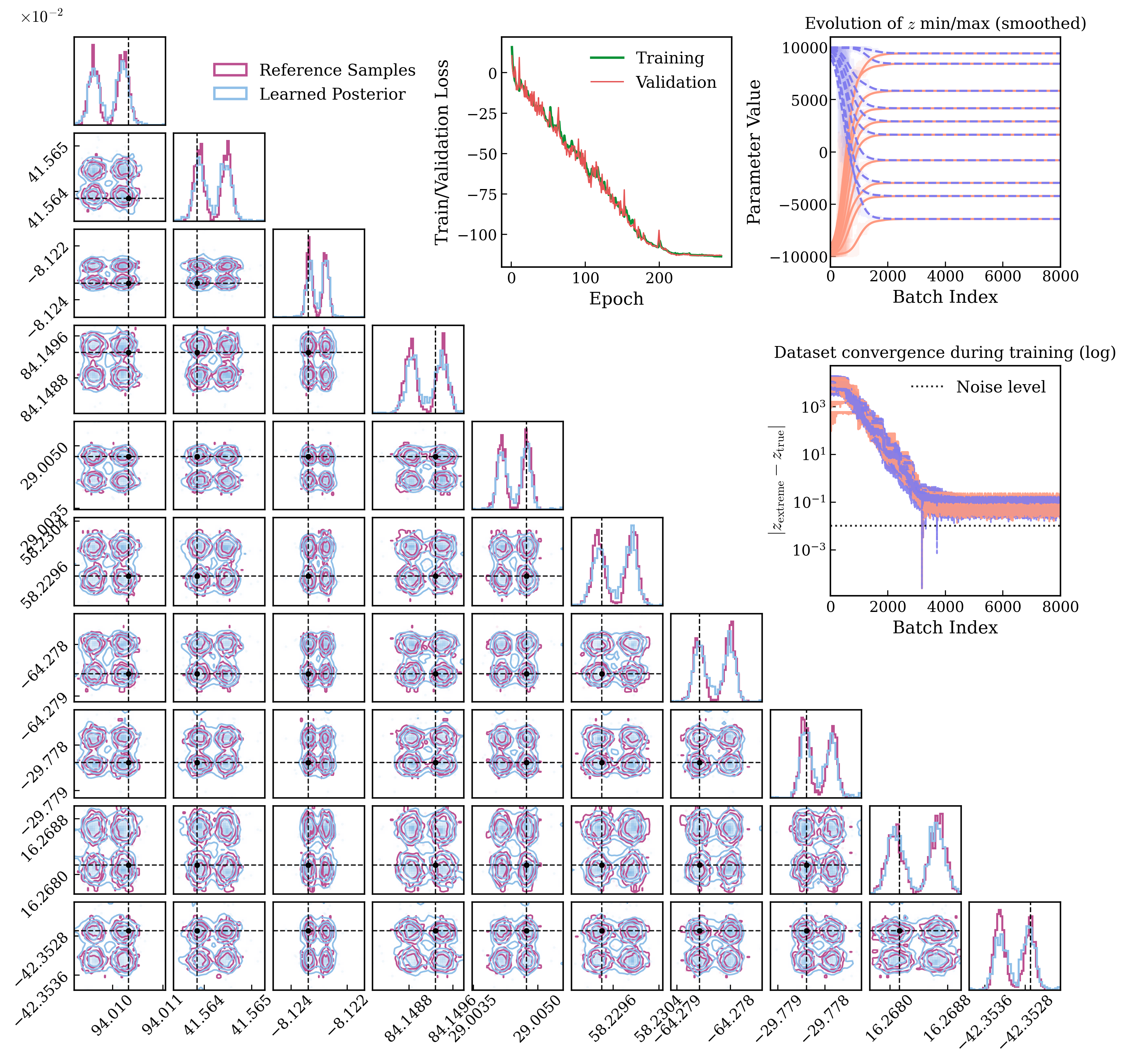}
    \caption{\textbf{10-D Bimodal Gaussian Mixture in DS-A:} Results for the synthetic benchmark described in Section~\ref{example1}.
    \emph{Left (corner plot):} Comparison between reference samples from the target posterior (pink) and samples from the learned posterior (blue). Black dashed vertical lines and markers indicate the two mode locations in each dimension.
    \emph{Middle upper (loss):} Training (green) and validation (red) losses versus epochs.
    \emph{Right upper (range):} Evolution of the per-dimension training parameter range, shown as smoothed \textbf{minimum} (orange, solid) and \textbf{maximum} (purple, dashed) across mini-batches.
    \emph{Right lower (convergence):} Dataset convergence during training, plotting $\lvert z_{\mathrm{extreme}}-z_{\mathrm{true}}\rvert$ per dimension (orange for minima, purple for maxima) on a log scale; the dotted black line marks the noise level. }
    \label{fig:exapmle1_OS_A}
\end{figure*}

In summary, the algorithm implements the following method choices.
\begin{itemize}
\item \texttt{loss}:
Train $q_{\phi}(\bz\mid\bx)$ with an importance–weighted NPE objective
$\displaystyle \mathcal{L}_\mathrm{DS-B}(\phi)= -(1/N)\sum w_i \log q_{\phi}(\bz_i\mid\bx_i)$,
where $w_i$ measures the relative generation probability of the parameter point from the proposal used to generate it, and the prior. This ensures that $q_\phi(\bz \mid \bx)$ directly converges to $p(\bz\mid\bx)$.

\item \texttt{proposal\_update}:
Set the proposal to the current posterior estimate given the target observation,
$\tilde p(\bz)\leftarrow q_{\phi}(\bz\mid \bx_{\mathrm{obs}})$.

\item \texttt{data\_update}:
Compute $c_B$ as the $\alpha_B$–quantile of $\{\log \tilde p(\bz_i)\}_{\bz_i\in\mathcal D_{\mathrm{live}}}$, remove points with $\log \tilde p^{\mathrm{new}}(\bz_i)<c_B$, re-simulate $\bz_{\mathrm{new}}\sim \tilde p$, and assign $w_{\mathrm{new}}=p(\bz_{\mathrm{new}})/\tilde p(\bz_{\mathrm{new}})$ (fixed thereafter).

\item \texttt{converged}: Stop once no new samples are deleted (and therefore generated) \emph{and} the validation loss has not decreased for a specified number of epochs.
\end{itemize}

\noindent The implementation details of DS-B within the meta-algorithm are laid out in Algorithm~\ref{alg:OS-B}.

\subsection{Other algorithm choices}

\noindent For completeness, we note that the sequential NPE family also includes SNPE-C (often referred to as APT)~\cite{Greenberg2019apt}. SNPE-C (APT) directly trains a conditional density estimator for the \emph{true} posterior without requiring post-hoc importance reweighting. As discussed in~\cite{Greenberg2019apt}, this has appealing theoretical properties: (i) the training objective is consistent under general proposal schedules, (ii) no explicit proposal–prior ratio correction is needed (reducing variance and hand-tuned weighting), and (iii) when the model class is sufficiently expressive, the learned posterior is asymptotically well-calibrated even under adaptive proposals. However, when applied to our high-dimensional, strongly multi-modal benchmark with extreme prior-to-posterior compression (Section~\ref{example1}), we found that our dynamic version of SNPE-C/APT did not train reliably in practice (due to instabilities and persistent degeneration of the estimator), and consequently failed to deliver usable posteriors. For this reason, in what follows, we focus our empirical analysis on DS-A and DS-B, for which we observed stable optimisation and accurate inference in the same setting. We do believe that alternative dynamic SBI schemes are possible, however, and encourage further algorithm development in future work.

\vspace*{6pt}
\noindent \textit{Code implementation details:} All code implementations can be found in the publicly available code packages within the \textbf{Dynamic SBI} GitHub organisation (\href{https://github.com/dynamic-sbi}{github.com/dynamic-sbi}), which serves as the main ecosystem that will scale going forward. Currently, DS-A examples are presented in \textbf{falcon-dsbi} (\href{https://github.com/dynamic-sbi/falcon-dsbi/tree/main}{github:falcon-dsbi}), based on the distributed dynamical SBI framework \textbf{Falcon} (\href{https://github.com/cweniger/falcon/tree/main}{github:falcon}), while \textbf{Chameleon} (\href{https://github.com/dynamic-sBI/chameleon-sbi}{github:chameleon-sbi}) provides a lightweight implementation of DS-B used for prototyping and physics applications. These repositories contain the configuration files and scripts required to reproduce all experiments reported in this work. The data relevant to this article are available publicly on our \href{https://zenodo.org/communities/dynamic-sbi/records?q=&l=list&p=1&s=10&sort=newest}{Zenodo} page.

\section{Benchmark Problems: Results and Performance}\label{sec:benchmarking}

\noindent In this section, we present a challenging benchmark analysis problem to demonstrate the performance of the dynamic SBI framework. In general, we are looking to test several properties, including i) how simulation-efficient the algorithm is, ii) how the quality of the posterior inference compares to amortised or traditional sequential methods, and iii) how fast the algorithm converges when the posterior-to-prior volume ratio is large.

\subsection{Example 1: High-precision Inference}
\label{example1}
\noindent To demonstrate the capability of our dynamic SBI framework (specifically the DS-A algorithm here) in high-dimensional, multi-modal settings, we consider a synthetic 10-dimensional Gaussian mixture model with two symmetric modes. The latent parameters $\mathbf{z} \in \mathbb{R}^{10}$ are assigned a broad uniform prior,
\begin{equation}
p(\mathbf{z}) = \mathcal{U}([-10^4,\,10^4]^{10}).
\end{equation}
Observations $\mathbf{x}$ are generated from a Gaussian likelihood centred around one of two equally weighted modes,
\begin{equation}
p(\mathbf{x} \mid \mathbf{z}) = \cfrac{1}{2}\mathcal{N}(\mathbf{x};\,\mathbf{z}+\boldsymbol{\mu}_+, \sigma^2 I) 
+ \cfrac{1}{2}\mathcal{N}(\mathbf{x};\,\mathbf{z}+\boldsymbol{\mu}_-, \sigma^2 I),
\end{equation}
where $\boldsymbol{\mu}_\pm = \pm 3\sigma \mathbf{1}$ denote the mode shifts in each dimension and $\sigma=10^{-2}$ defines the scale of observational noise. This corresponds to a prior volume approximately $10^{60}$ times larger than the true posterior support. A single observation $\mathbf{x}_0$ is drawn from this generative model and used as input for inference.

The details of the normalising flow architectures, the embedding networks, and the complete training setup, including optimiser configuration and dynamic sample management, are provided in Appendix~\ref{appendix:os_a_details}. Here, we focus on the key outcomes that highlight the efficiency and accuracy of the proposed dynamic SBI method.

\vspace*{6pt}
\noindent \textbf{Posterior Recovery and Dataset Focusing:} The main results concerning the convergence properties of the DS-A algorithm are shown in Fig.~\ref{fig:exapmle1_OS_A}. In particular, we highlight the convergence and general agreement with reference samples from the true posterior, the dynamics of the training and validation loss, as well as the evolution and convergence of the dynamical dataset. Overall, these results demonstrate the ability of the dynamic SBI algorithm to efficiently converge for problems with a very large posterior-to-prior ratio.

In more detail, the corner plot (lower left panels) indicates the close agreement between reference samples from the analytic posterior (pink) and samples from the learned posterior (blue); black dashed vertical lines and point markers denote the true parameters that produce the observation. Quantitatively, we computed a Jensen-Shannon Divergence of $\mathrm{JSD}\approx 0.073$ between the reference samples and the samples from the DS-A algorithm. In addition, the optimisation dynamics are stable: both training and validation losses decrease smoothly to their converged value without indications of overfitting (middle upper panel of Fig.~\ref{fig:exapmle1_OS_A}). 

\begin{figure*}[t]
    \centering
    \includegraphics[width=\linewidth]{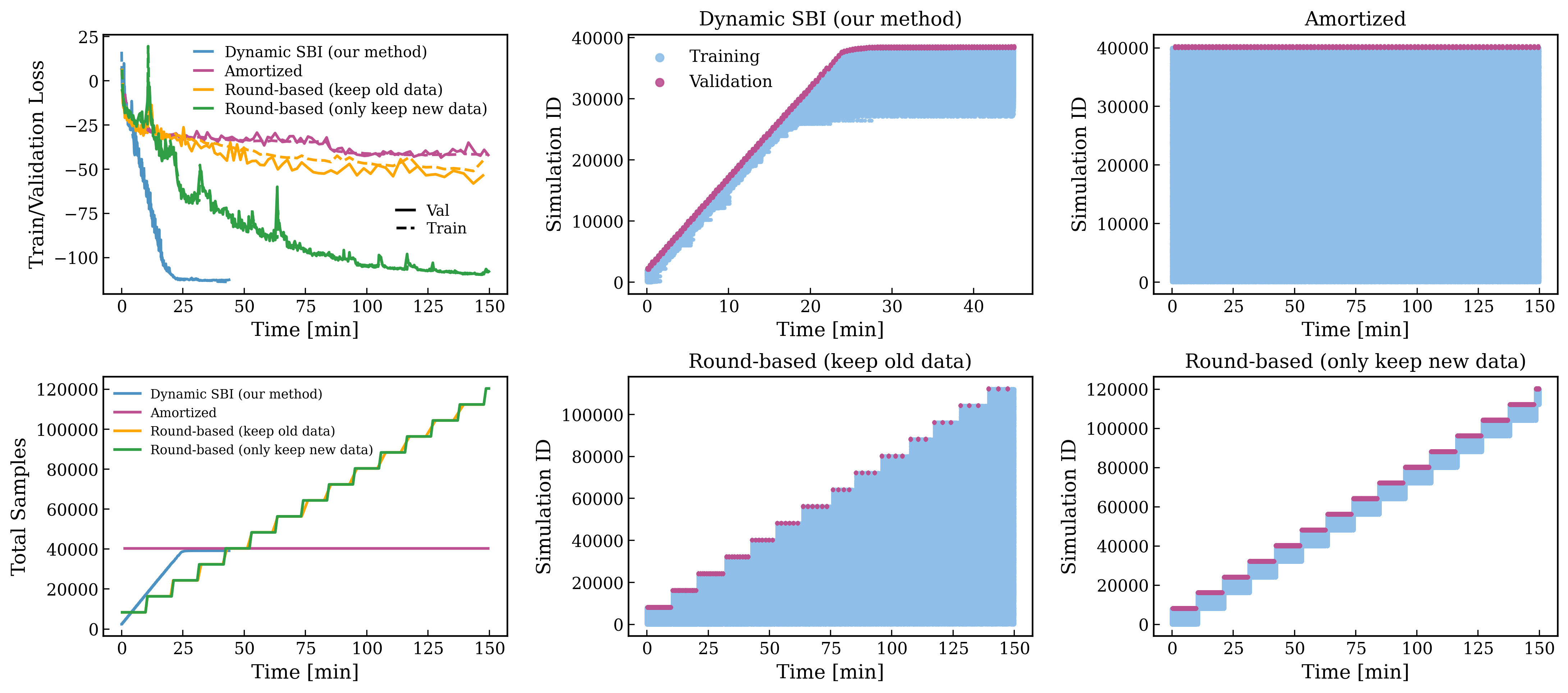}
    \caption{
        \textbf{Comparison of DS-A, Round-based, and amortised methods.} \emph{Left upper:} Training (dashed) and validation (solid) losses versus \emph{elapsed} time (minutes from the start) for four schemes: Dynamic SBI (blue), Amortised (pink), Round-based (keep old data; orange), and Round-based (only keep new data; green). \emph{Left lower:} Cumulative number of simulations versus elapsed time for the same four schemes. \emph{Right four panels:} Evolution of simulation identifiers over elapsed time for each scheme separately. In these panels, blue dots denote training simulations and magenta dots denote validation simulations.}
    \label{fig:round_free_amortized_compare_OS_A}
\end{figure*}

Finally, we show the convergence of the dynamical dataset itself in the upper right panels by studying the range of the per-parameter maximum/minimum samples in the dataset as a function of training progress. In this regard, the upper right panel shows smoothed minima (orange, solid) and maxima (purple, dashed) across training mini-batches. We observe that after approximately $2{,}000$ batches, these envelopes have contracted to the vicinity of the true posterior support. This ultimately demonstrates that the initially uninformative prior domain $[-10^{4},\,10^{4}]$ in each parameter dimension collapses within the first few thousand mini-batches, concentrating the live set $\mathcal{D}_{\mathrm{live}}$ around the ground-truth posterior support and thereby enabling efficient exploration of the parameter space. To facilitate clearer visualisation of this late-stage behaviour, the right lower panel presents a logarithmic version of this convergence. In particular, for each parameter $z_i$, we plot the behavior of $\mid z_i^{\max}-z_i^{\mathrm{true}}\mid$ and $\mid z_i^{\min}-z_i^{\mathrm{true}}\mid$ where $z_i^\mathrm{min, max}$ is the minimum/maximum value of $z_i$ currently in the dataset. A dotted horizontal line indicates the statistical noise level $\sigma =10^{-2}$, highlighting the fact that the live dataset stabilises once the support of the posterior has been covered.

\vspace*{6pt}
\noindent \textbf{Comparison: DS-A vs. Round-Based Variants and Amortised Baseline} To further illustrate the advantages of dynamic SBI, we compare DS-A against two variants that resemble round-based sequential algorithms, as well as an amortised baseline on the same $10$D bimodal problem. On the same hardware\footnote{Experiments were run on a single NVIDIA GeForce RTX 3080~Ti (12\,GB), driver~570.172.08, CUDA~12.8.}, all methods are given a comparable wall-clock budget ($\sim 150\text{ min}$), from which we can compare convergence, simulation budget, and training stability as compared to our benchmark run above. The main results here are shown in Fig.~\ref{fig:round_free_amortized_compare_OS_A}, which highlights the comparison of the training losses over time (left upper panel), as well as the different simulation budgets required by each version of the algorithm (remaining panels).

Our baseline comparison point is the same as presented in Fig.~\ref{fig:exapmle1_OS_A} using the DS-A algorithm. In this implementation, we continuously curate the training set given the current statue of the posterior estimator: every $\sim 15 \text{ s}$ we replenish the simulation pool to maintain at least $2048$ live simulations and prune samples with low posterior mass under the current posterior network (as defined by the threshold in Eq.~\eqref{eq:rthreshold}). This is shown in the upper middle panel of Fig.~\ref{fig:round_free_amortized_compare_OS_A}, where we observe that the training set converges after approximately 20 minutes, with no further new simulations generated. To achieve the results presented above, DS-A takes approximately $\sim 25\text{ min}$ of wall-time and a total of $\sim 40,000$ simulations. We take these timing and simulation budget numbers as our reference points for the other methods.

For the first comparison point, we train an amortised method once on a fixed set of $40,000$ simulations. We see from the upper left panel of Fig.~\ref{fig:round_free_amortized_compare_OS_A} that this is not a sufficiently large simulation budget to resolve the details of the high-precision posterior under study (indicated by the quick saturation of the loss curve). The takeaway message from this comparison is that in this problem, either a) far more simulations are required to obtain an amortised estimator, or b) a sequential method is required. This sets an upper envelope for the algorithm performance.

In our second comparison (which we call \emph{Round-based (keep old data)}), new batches of 8,000 simulations from the current proposal are appended every $\sim 10 \text{min}$ without discarding old data; the total sample count grows in steps and the loss improves after each addition but remains dominated by stale, overly broad samples from earlier rounds. We observe that, despite a large total number of simulations (over 120,000), the quality of inference is not significantly improved over the amortised method. Again, the takeaway here, as compared to the DS-A algorithm, is that maintaining all samples in the dataset, even when they are highly disfavored, is actually extremely inefficient as far as posterior recovery is concerned.

Our final comparison point is a more standard implementation of round-based sequential SBI (we call it \emph{Round-based (only keep new data)} here to distinguish it from the last case study). Here, every $\sim 10 \text{min}$, new simulations are sampled from the current status of the proposal and all old simulations are removed. This mitigates the issues associated with training on all samples seen above, and we observe from Fig.~\ref{fig:round_free_amortized_compare_OS_A} that we are indeed able to converge to the high-precision result obtained using DS-A. However, we also observe that this takes the full 150 minutes of wall-time and a factor of three (120,000 in total) more simulations. We also note that this method exhibits transient loss spikes at round boundaries, consistent with abrupt shifts of the training distribution before the network quickly re-adapts.

To conclude this results section, we see that for equal budgets, our new algorithm (DS-A) delivers the fastest, most efficient, and smoothest posterior recovery. This ablation study indicates the core takeaways that: (i) short resampling intervals and (ii) active pruning of low-posterior samples are the key drivers of efficiency. Together, they continually steer simulations toward informative regions while avoiding the accumulation of uninformative data. These properties indicate that dynamic SBI can be particularly effective in high-dimensional, multi-modal settings with extreme prior-to-posterior compression, achieving accurate posteriors with fewer simulations and markedly faster convergence than amortised and round-based alternatives.

\subsection{The continuum limit of sequential SBI}

\noindent One interesting framing of the algorithms presented in this work, given these comparison results, is the concrete connection between dynamic SBI and typical round-based approaches. In some ways, dynamic SBI can be seen as the practical limit of round-based approaches where the number of rounds $R \rightarrow \infty$ and the length of each round (measured in, say, training epochs) $L_R \rightarrow 0$. As mentioned above, the utility of this limit lies in the assumption that there is an inherent redundancy in most sequential round-based approaches, where simulation data is still used for a significant number of training steps even if it would already be highly disfavored by the current estimator. Taking the length of each round to zero is then the statement that this deletion criterion can be checked continuously. Of course, there are practical limitations\footnote{In the current implementation, we do not monitor the dataset updates at a cadence faster than once per epoch, but this is not an algorithmic limitation, and we will revisit this aspect in future updates to the code.} regarding how closely this algorithmic limit can be reached in principle concerning: a) how quickly one can check the dataset $\mathcal{D}_{\mathrm{live}}$ for possible deletions, b) how quickly one can sample from the proposal $\tilde{p}(\bz)$, and c) how fast one can then regenerate the data using the computational simulator $p(\bx \mid \bz)$. Ultimately, the dominant effect will vary case-by-case and depends broadly on the ratios between: the time to sample, $t_\mathrm{sample}$, from $q_\phi(\bz \mid \bxobs)$ (which motivated our choice of NPE as a base algorithm); the typical time associated to taking a single training step, $t_\mathrm{step}$ (which depends on network complexity, capacity, and hardware); and finally the time to run the simulator $\bx \sim p(\bx \mid \bz)$, $t_\mathrm{sim}$. Ultimately, this discussion highlights the fact that the specific resampling time ($10 \mathrm{\ mins})$ chosen above for comparison was not the driver of the result, and we can always expect the corresponding round-based method to lie within the dynamic SBI/amortised envelope that we presented in the upper panel of Fig.~\ref{fig:round_free_amortized_compare_OS_A}.

\section{Full-scale Examples: Gravitational Waves and Strong Lensing}\label{sec:physics}

\noindent After benchmarking the performance of our dynamic framework, we proceed by demonstrating the capability to address problems in astrophysics. In particular, we study two compelling data analysis challenges in the fields of gravitational waves (GW) and strong lensing.

\begin{figure*}[t]
    \centering
    \includegraphics[width=\linewidth]{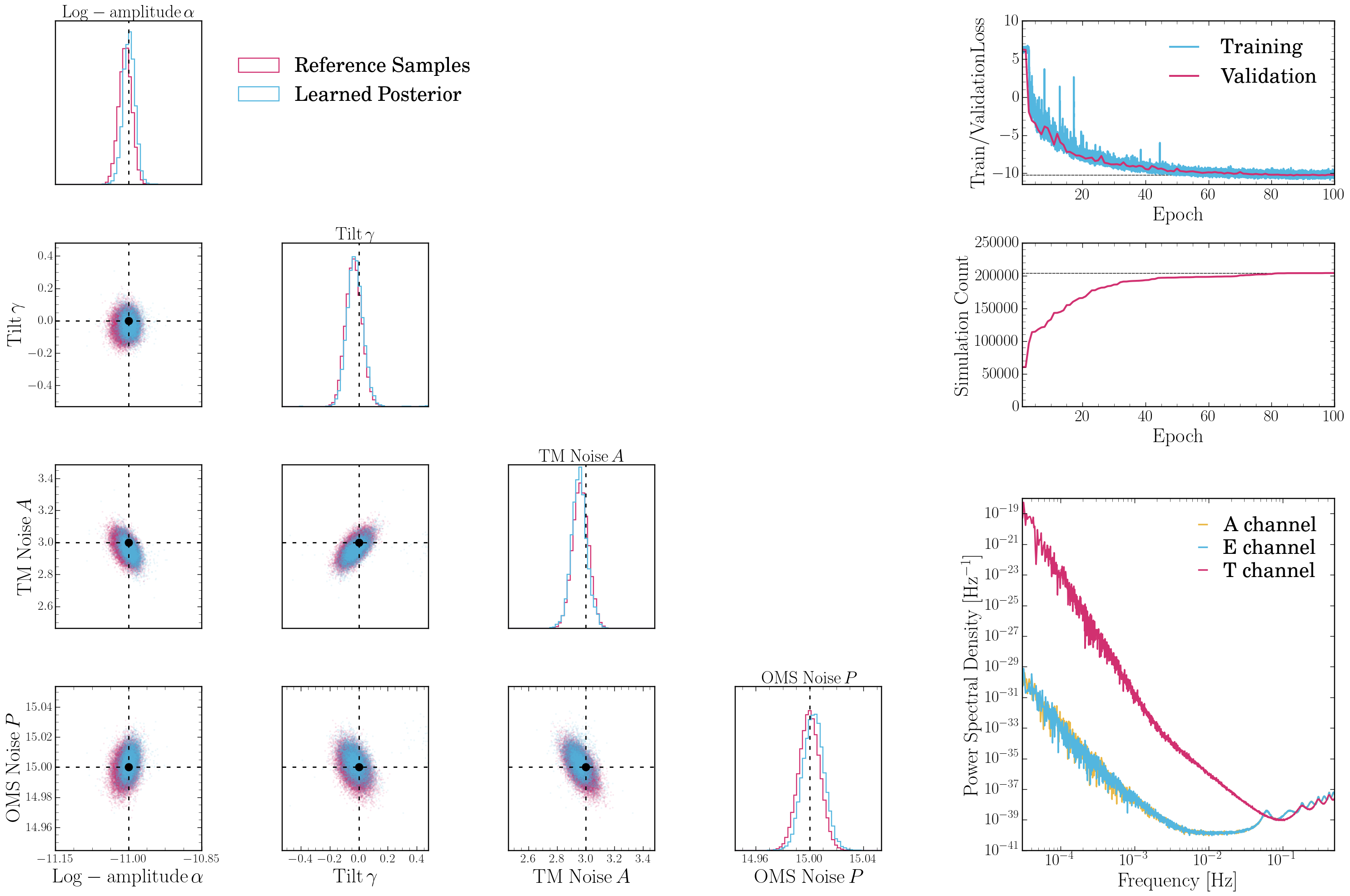}
    \caption{\textbf{Stochastic Gravitational Wave Background:} Analysis results for the SGWB case study presented in Section~\ref{sec:physics}. \emph{Left corner plot:} Comparison between the reference samples (pink) and samples from the learned posterior (blue) for the benchmark case study described in the main text. The true values of the injection parameters are shown with dotted black lines and dots. \emph{Right upper:} The training (blue) and validation (pink) losses as a function of the number of training epochs. \emph{Right middle:} The total simulation count (across training and validation) used in the training run. \emph{Right lower:} The observation data analysed in this case study from the three different (A, E, and T) data channels.}\vspace*{-10pt}
    \label{fig:sgwb}
\end{figure*}

\subsection{Reconstructing the Stochastic Gravitational Wave Background}\label{subsec:sgwb}

\noindent One of the most important developments in the next generation of GW astrophysics will be the launch of the Laser Interferometer Space Antenna (LISA)~\cite{LISA:2017pwj}. LISA is a space-based GW observatory targeting the milli-Hertz (mHz) frequency range of the GW spectrum. It promises to revolutionise our understanding of a wide range of astrophysical and cosmological phenomena (see Ref.~\cite{Colpi:2024xhw} for a detailed review), including: the properties and distribution of supermassive black holes in the Universe; binary systems in the Milky Way; and even environments in the disks and dark matter haloes of host galaxies~\cite{Cole:2022yzw}. One crucial source class beyond these resolvable, deterministic signals, are so-called stochastic GW backgrounds (SGWBs). These stochastic sources originate from the incoherent superposition of a large number of GW sources (astrophysical signals) or from violent processes in the early Universe (cosmological signals). Concerning astrophysical SGWBs, the loudest signals are expected to originate from white dwarves, both galactic~\cite{Bender:1997hs,Barack:2004wc,Karnesis:2021tsh,Toubiana:2024qxc} and extragalactic~\cite{Staelens:2023xjn,Hofman:2024xar,
Boileau:2025jkv}, and extragalactic stellar-origin binary black holes (with a minor contribution from binary neutron stars)~\cite{Cusin:2019jhg,Cusin:2019jpv,Perigois:2020ymr,Lewicki:2021kmu,Babak:2023lro,Lehoucq:2023zlt}. Cosmological sources include topological defects~\cite{Blanco-Pillado:2024aca}, first-order phase transitions~\cite{Caprini:2018mtu}, inflation~\cite{LISACosmologyWorkingGroup:2024hsc}, and scalar-induced GWs~\cite{LISACosmologyWorkingGroup:2025vdz}, see, \emph{e.g.},~\cite{Caprini:2018mtu,LISACosmologyWorkingGroup:2022jok}. Therefore, the confident recovery and reconstruction of SGWB components at LISA is an important data analysis challenge to solve.

There are a number of classical sampling-based approaches to SGWB reconstruction (see, \emph{e.g.},~\cite{Cornish:2005qw,Flauger:2020qyi,Baghi:2023qnq,Santini:2025iuj,Pozzoli:2023lgz}), as well as multiple SBI pipelines developed to carry out Bayesian inference on both astrophysical and cosmological SGWB sources~\cite{Alvey:2023npw,Dimitriou:2023knw,Alvey:2024uoc}. In this work, we aim to demonstrate the capacity of the dynamic framework to handle the analysis at the scale of the full mission. As such, we take the example presented in Ref.~\cite{Alvey:2024uoc}, which looked to analyse the scenario in which instrumental noise in LISA was not stationary over the mission duration. Here, we demonstrate that we can use the dynamic framework to carry out the benchmark analysis task at the heart of that work: constructing posterior estimators for the parameters of the instrumental noise model (denoted $A$ and $P$ below) and the parameters of a power-law template (the log-amplitude $\alpha$ and tilt $\gamma$) across a single 11.5 day segment.\footnote{Note that typically for SGWB analyses, the data is averaged and coarse-grained over the approximately 100 segments (amounting to about 3 effective years of data), see, \emph{e.g.},~\cite{LISA_performance}. As such, the analysis presented here still reflects the typical level of complexity for this class of inference problem.} To ensure a like-for-like comparison, we follow as closely as possible the inference and simulation setup described in Ref.~\cite{Alvey:2024uoc}. In particular, we refer the reader to the following parts of this reference for further details: Section II.B (LISA noise model), Section II.C (Instrumental Response), Section II.D (Power-law signal model), Section III.A (Fiducial parameter choices), Section IV.A (Coarse-graining strategy and Data Generation), and Section IV.B (Network architecture). 

In this work, we take similarly wide priors on the noise and signal parameters given by $p(A, P, \alpha, \gamma) = \mathrm{U}_A(0, 6) \times \mathrm{U}_P(0, 30) \times \mathrm{U}_\alpha(-20, -5) \times \mathrm{U}_\gamma(-5, 5)$, and generate a coarse-grained resampling dataset of size $10^5$ (see Section IV.A of Ref.~\cite{Alvey:2024uoc}). To generate the results in Fig.~\ref{fig:sgwb}, we used a training dataset of size $50,000$, a validation dataset with size $10,000$, an initial learning rate of $5 \times 10^{-4}$, and a learning scheduler that consisted of a linear warm-up phase of 5 epochs, following by a cosine annealing of 95 steps for a total of 100 epochs of training.\footnote{In terms of computational resources, the network was trained on an NVIDIA RTX 4500 Ada Generation graphics card, and the simulations carried out on 32 Intel(R) Core(TM) i9-14900K CPUs.} In addition, as well as the compression network described in Ref.~\cite{Alvey:2024uoc}, we train an NSF~\cite{Durkan2019nsf} as implemented in the \texttt{sbi} package~\cite{TejeroCantero2020sbi} with $z$-scoring on the input data turned off, using the DS-B algorithm described in Section~\ref{sec:online}. The full implementation of the simulator, network architectures, and training steps is provided in the examples folder of the \href{https://github.com/dynamic-sbi/chameleon-sbi}{github:chameleon-sbi} library.

\subsection{Strong Gravitational Lensing}\label{subsec:lensing}

\begin{figure*}[t]
    \centering
    \includegraphics[width=\linewidth]{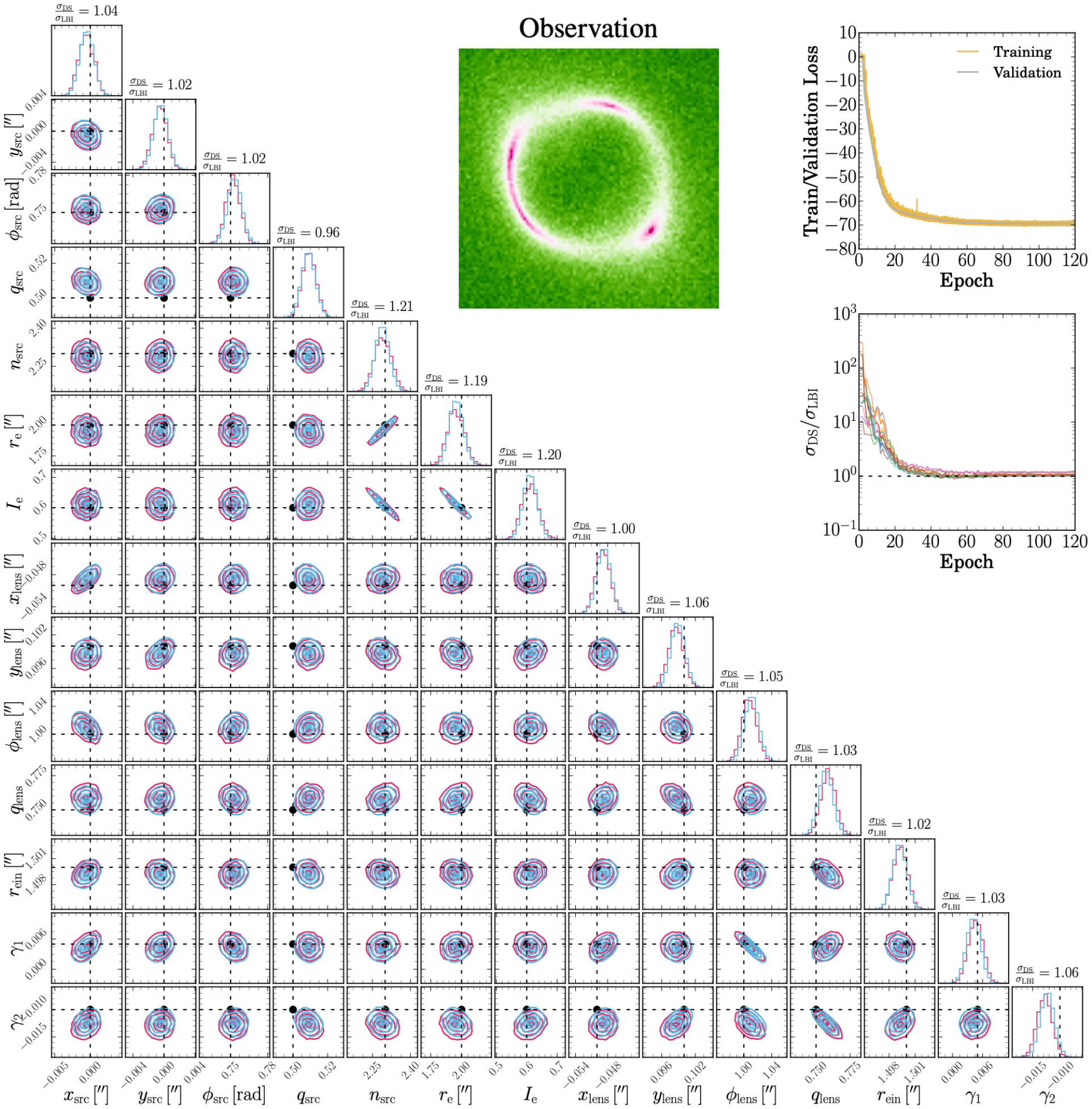}
    \caption{\textbf{Strong Gravitational Lensing:} Analysis and inference results for the strong gravitational lensing example presented in Section~\ref{subsec:lensing}. \emph{Main corner plot:} Comparison between LBI (blue contours, overlaid) and DS (pink contours) inference results for all model parameters. Black dashed lines and black points indicate the true values for reconstruction (see Table~\ref{tab:lensing-params} for parameter definitions, values, and priors). On top of each 1D marginal histogram, we report the ratio of the standard deviations between the final DS posterior and the LBI posterior, showing close consistency across all parameters. \emph{Right upper inset:} Training (yellow) and validation (grey) loss curves for the DS-B method as a function of training epoch. \emph{Right lower inset:} Evolution of the ratio between the proposal standard deviation and that of the likelihood-based results, demonstrating a rapid reduction in variance and convergence toward the target posterior distribution. \emph{Middle inset:} Target observation under consideration.}
    \label{fig:strong_lensing}
\end{figure*}

\noindent To test our algorithm in a more realistic and high-dimensional setting, we consider a simplified astrophysics problem: strong gravitational lensing parameter estimation.  Galaxy-galaxy strong gravitational lensing occurs when the mass of an intervening galaxy distorts the paths of light rays from a background source before they reach a telescope~\cite{lensing}. This phenomenon can produce highly distorted, ring-shaped images, often with multiple copies of the source galaxy. By carefully analysing the relationships between these multiple images, one can precisely characterise the properties of both the lens and the source.
In recent years, a sequential SBI approach has been well-motivated for this type of analysis and has been proven successful~\cite{Coogan:2022cky, Anau_Montel_2022, Wagner-Carena:2024axc}. In this example, we test our dynamic approach against a standard likelihood-based inference (LBI) technique, nested sampling, to verify its accuracy on this simplified astrophysical problem.

We generate simulated strong lensing observations using the simulator adopted in~\cite{coogan2020targeted, coogan2022walks, Anau_Montel_2022}. Our model of the lensing system consists of the lens mass and source light. The surface brightness of the background source is modelled using a S\'ersic profile~\cite{Sersic_1963}, parameterised by seven variables: central position, position angle, axis ratio, S\'ersic index, effective radius, and surface intensity. For the main lens mass distribution, we adopt a singular power-law ellipsoid (SPLE) model~\cite{Suyu_2009}, which has six parameters: position, position angle, axis ratio, slope, and Einstein radius. We also include two additional parameters to model the external shear. The priors for these parameters and the fiducial values used for our mock observation are shown in Tab.~\ref{tab:lensing-params}.
Based on this model, we generate 100 x 100 pixel images with a resolution of 0.05$''$ per pixel, for a total field of view of 5$''$ x 5$''$. We assume a simple instrumental noise model, adding Gaussian noise to each pixel drawn from a distribution with a standard deviation of $\sigma_n = 1.0$, measured in units of flux. For our simulations, we set the lens redshift to $z_\mathrm{lens} = 0.9$ and $z_\mathrm{src} = 2.0$.

For the benchmark results presented in Fig.~\ref{fig:strong_lensing}, we carry out two analyses. The first is a likelihood-based sampling run where we use the \texttt{nautilus} software package~\cite{nautilus} to generate an effective ``ground truth'' set of posterior samples that we can compare our dynamic implementation with. To achieve this, we implement a Gaussian likelihood in pixel space assuming that the instrumental noise in each pixel is uncorrelated and has a variance of $\sigma_n^2 = 1$ as measured in units of flux. For the run displayed in Fig.~\ref{fig:strong_lensing}, we used 2000 live points. We also explicitly checked that re-running the analysis with these settings resulted in identical results at the level of the posterior. In total, this generated a set of posterior samples with $n_\mathrm{eff} \simeq 2.8 \times 10^4$ effective samples.

For the dynamic analysis, we again used the DS-B algorithm as a benchmark. To allow a direct comparison with the SGWB example in the previous section, we ran with a training dataset of size $50,000$ and a validation dataset with size $10,000$, and an identical learning rate scheduler setup (\emph{i.e.}, an initial warm-up phase of 5 epochs followed by 115 epochs with a cosine annealing schedule). In this case, we ran with a maximum learning rate of $10^{-3}$ and a minimum learning rate of $10^{-8}$. Similarly, we used the same NSF architecture described in Section~\ref{subsec:sgwb}, with an embedding network that processes the input lensing image identical to the one described in Table III of Ref.~\cite{AnauMontel:2023stj}. This embedding network broadly consists of a dynamic normalisation layer followed by a set of standard convolutional layers. As in the SGWB example, the full implementation and code examples can be found in the examples folder of the \href{https://github.com/dynamic-sbi/chameleon-sbi}{github:chameleon-sbi} library.

The results for this analysis are presented in Fig.~\ref{fig:strong_lensing}. In the main corner plot, we show that we can achieve close agreement with the likelihood-based sampling results. Note that we obtained this strong agreement for posteriors obtained from a prior volume compression of the order of a factor $10^{31}$ in this 14-dimensional parameter space. The bottom-right panel shows the evolution for each parameter of the ratio between the proposal’s standard deviation and that of the likelihood-based results, indicating a significant reduction in data variance relative to the initial prior range. Finally, we demonstrate the smooth evolution of the training process as the dataset evolves via the training and validation curves.

\section{Outlook and Conclusions}\label{sec:conclusions}

\noindent In this work, we have introduced a ``dynamic'' framework for sequential SBI. The core novelty of our approach lies in treating simulation and network training as parallel, asynchronous processes communicating through a shared, dynamic dataset (see Fig.~\ref{fig:summary} and Algorithm~\ref{alg:NPE}). This ``live'' dataset is continuously pruned of uninformative samples and replenished with new simulations in high-probability regions for the observation being targeted. Ultimately, the goal of this process is to ensure that computational resources are always focused where they are most useful. We demonstrated the power of this approach on a challenging 10-dimensional benchmark (see Section~\ref{sec:benchmarking}) with a prior-to-posterior volume compression factor of $\sim 10^{60}$, showing rapid convergence and accurate posterior recovery. Furthermore, we validated our framework on two astrophysical analysis problems (see Section~\ref{sec:physics}): reconstructing the SGWB and analysing strong gravitational lensing systems, demonstrating its capacity for handling scientific statistical inference problems.

\subsection{Development Outlook}

\noindent Looking forward, a significant opportunity lies in extending the dynamic framework to large-scale, multi-component models. The forthcoming LISA mission provides a prime astrophysical example with its ``global fit'' challenge~\cite{Cornish:2005qw}, which requires the simultaneous characterisation of tens of thousands of galactic binaries, multiple massive black hole mergers, and various stochastic/instrumental backgrounds, see, \emph{e.g.}, Refs.~\cite{Littenberg:2023xpl,Strub:2024kbe,Katz:2024oqg,Deng:2025wgk} for existing likelihood-based implementations. The sheer scale and variable dimensionality of this problem would pose a severe challenge for rigid, round-based sequential methods. Our framework is naturally suited to this task if the dynamic dataset is allowed to interact with multiple inference ``nodes''. We leave this as a high-priority area for algorithm development (to handle, \emph{e.g.}, the fact that the dataset can be altered by more than one proposal) and example benchmarking. There are also natural connections to efficient distributed computing frameworks in this large-model framing.

More broadly, the dynamic SBI framework is designed to alleviate issues with simulation and/or network training bottlenecks. In particular, as we saw in Section~\ref{sec:benchmarking}, there are two limits to this scenario: if simulation cost dominates, then dynamic SBI can help by more efficiently zooming in to high-probability regions of the relevant parameter space. On the other hand, if training costs dominate, then the dynamic framework can assist by updating the dataset as soon as there is evidence that simulations are disfavoured. This ensures that as little time as possible is spent training on simulations that are uninformative for the observation at hand. There are a number of key astrophysical applications that fit these criteria. For example, the analysis of Extreme Mass Ratio Inspirals (EMRIs), whose gravitational waveforms are notoriously expensive to compute, has been shown to render traditional sampling techniques extremely computationally intensive (see, \emph{e.g.}, Refs.~\cite{Gair:2004iv,Chua:2021aah,Cornish:2008zd}). In addition, the data can often be high-dimensional, leading to large networks and high training costs in the SBI context. Another prominent example lies in modern cosmology, where inferring parameters from large-scale structure observables often relies on computationally demanding N-body or hydrodynamical simulations. Whether analysing galaxy clustering, weak lensing shear maps, or the Lyman-alpha forest, the cost per simulation is extremely high, making an approach that minimises redundant simulations essential for a tractable analysis (see, \emph{e.g.}, Refs.~\cite{Alsing2019,Jeffrey2024,Lemos2022,Cranmer2020}). 

A final powerful extension of our framework could be its application to hierarchical inference, where the goal is to constrain both the properties of individual events and the population-level distributions from which they are drawn (see, \emph{e.g.}, Ref.~\cite{AnauMontel:2022ppb}). This is essential for turning catalogues of observations, such as GW events (\emph{e.g.},~\cite{KAGRA:2021duu}) or strong lensing systems (\emph{e.g.},~\cite{Bolton:2008xf}), into fundamental knowledge about astrophysics, like the black hole mass function or the properties of dark matter subhalos (\emph{e.g.},~\cite{KAGRA:2021duu,Vegetti_2012}). The primary technical hurdle in this setting would be to extend the dynamic proposal and live dataset framework to the nested structure of hierarchical models. This would require developing a proposal mechanism that can efficiently explore the joint space of population and event-level parameters, as well as a principled criterion for pruning and replenishing simulations of entire catalogues. In this setting, it would also be interesting to explore integration of dynamic SBI with multi-fidelity simulation (see, \emph{e.g.},~\cite{Krouglova:2025aaa}). In this more advanced setup, the algorithm could be extended to dynamically request either fast, low-fidelity simulations for broad exploration or expensive, high-fidelity ones for final, precise inference. Such a strategy would represent a compromise in computational efficiency, allocating resources not just based on parameter space location but also on the required simulation accuracy at each stage of the analysis.

\subsection{Key Conclusions}

\noindent We have presented a flexible and computationally efficient paradigm for sequential SBI that moves beyond the static, round-based structure of existing methods. By enabling the dataset and proposal to evolve dynamically and asynchronously, our dynamic framework significantly reduces the total inference cost for a range of problems. The primary advantages of our dynamic SBI framework can be summarised as follows:
\begin{itemize}
    \item \emph{Asynchronous, Sequential Inference:} Dynamic SBI eliminates the rigid structure of sequential rounds, allowing for continuous and parallel adaptation of both the training dataset and the parameter proposal distributions. This is summarised in Fig.~\ref{fig:summary} and in Algorithm~\ref{alg:NPE}.
    \item \emph{Algorithm Development:} We presented two versions of the dynamic SBI algorithm: DS-A and DS-B (see Section~\ref{sec:online}). These broadly follow the taxonomy of the current suite of SNPE algorithms~\cite{Papamakarios2016npe,Lueckmann2017flexible,Greenberg2019apt}. The core innovation in this regard that enabled the dynamical aspect of the algorithms were two novel proposal/rejection schemes. We expect that further iterations and significant improvements to these baseline algorithms are possible, and are a key area for future work.
    \item \emph{Cost Efficiency:} By rapidly discarding improbable simulations and replacing them with targeted new ones, the framework minimises wasted computational effort in terms of both simulator cost and training time. We benchmarked this efficiency gain by comparing to sequential round-based SBI algorithms, as well as an amortised setup, in Fig.~\ref{fig:round_free_amortized_compare_OS_A}.
    \item \emph{Robust High-Dimensional Performance:} The methodology delivers accurate and well-calibrated posterior estimates for complex, multi-modal problems requiring extreme compression from prior to posterior, as demonstrated in our benchmark example and physics applications (see Sections~\ref{sec:benchmarking} and~\ref{sec:physics}).
\end{itemize}
We believe that dynamic SBI should be the preferred method when the cost of simulation is high and the overhead associated with redundant training and/or round-based analysis structures becomes a significant bottleneck. This work ultimately sets the foundation for a new class of sequential SBI algorithms.

\section*{Acknowledgments}

\noindent We would like to thank David Yallup for helpful conversations about connections to Bayesian computation, and Fabian Schmidt for helpful comments on the draft of the manuscript. JA is supported by a fellowship from the Kavli Foundation. The work of MP is supported by the Comunidad de Madrid under the Programa de Atracción de Talento Investigador with number 2024-T1TEC-3134. MP and JA acknowledge the hospitality of Imperial College London, which provided office space during some parts of this project. The work of CW was supported by a project that has received funding from the European Research Council (ERC) under the European Union’s Horizon 2020 research and innovation program (Grant agreement No. 864035 – UnDark). HL is supported by a fellowship from the China Scholarship Council (CSC).

\bibliography{main}
\appendix

\begin{algorithm*}
\caption{\textbf{Dynamic Simulation-Based Inference–A (DS-A)} }
\label{alg:OS-A}
\begin{algorithmic}[1]

\State \textbf{Inputs:}
\State \hspace{\algorithmicindent} Prior $p(\bz)$,\; simulator $p(\bx\mid\bz)$,\; observation $\bx_0$
\State \hspace{\algorithmicindent} Initial dataset $\mathcal{D}_\mathrm{live}=\{(\bz_i,\bx_i)\}_{i=1}^{N_0}$ with $\bz_i\!\sim\! p(\bz)$,\; $\bx_i\!\sim\!p(\bx\mid\bz_i)$
\State \textbf{Outputs:} Trained conditional flow $q_\phi(\bz\mid\bx)$ (proposal-biased posterior) 
and unconditional flow $q_\psi(\bz)$ (auxiliary)

\Statex
\State \textbf{Initialise:} build $q_\phi(\bz\mid\bx)$ and $q_\psi(\bz)$, fix tempering $\gamma\!\in\!(0,1]$, choose discard threshold $\tau$ (\emph{e.g.},\ $\tau=-20$), use $J$ to represent the number of new samples.

\Statex
\State \textbf{Main dynamic loop}
\While{\texttt{not converged}}
  \State \textbf{(Train posterior)}\quad minimise $\mathcal{L}_{\mathrm{post}}(\phi) = -\cfrac{1}{B}\!\sum_{(\bz,\bx)\in\text{batch}}\log q_\phi(\bz\mid\bx)$
  \State \textbf{(Train auxiliary)}\quad minimise $\mathcal{L}_{\mathrm{aux}}(\psi) = -\cfrac{1}{B}\!\sum_{(\bz,\bx)\in\text{batch}}\log q_\psi(\bz)$
  \State \textbf{(Proposal update)}\quad $\tilde p(\bz)\leftarrow \texttt{proposal\_update}\big[q_\phi(\bz\mid\bx_0)\big]$
  \State \textbf{(Data update)}\quad $\mathcal{D}_\mathrm{live}\leftarrow \texttt{data\_update}\big[\mathcal{D}_\mathrm{live},\,\bx_0,\,\tilde p(\bz),\,q_\phi,\,\gamma,\,J,\,\tau\big]$
\EndWhile

\Statex
\State \textbf{function} \texttt{data\_update}$\big[\mathcal{D}_\mathrm{live},\,\bx_0,\,\tilde p(\bz),\,q_\phi,\,\gamma,\,J,\,\tau\big]$
\State \hspace{\algorithmicindent}remove $(\bz_i,\bx_i)$, if $r_i=\log q_\phi(\bz_i\mid\bx_0)-\log q'_\phi(\bz_i | \bx_0) <\tau$
\State \hspace{\algorithmicindent}draw $K$ candidates $\{\bz_k\}\!\sim\! q_\phi(\bz\mid\bx_0)$;
\State \hspace{\algorithmicindent}set $\ell_k^{\mathrm{post}}=\log q_\phi(\bz_k\mid\bx_0)$; 
\State \hspace{\algorithmicindent}$\text{mask}_k\!=\!100$ if $\bz_k$ out of prior, else $0$
\State \hspace{\algorithmicindent}define $\log w_k=-\cfrac{1}{1+\gamma}\,\ell_k^{\mathrm{post}}-\text{mask}_k$
\State \hspace{\algorithmicindent}normalise and resample $J$ indices with replacement from $\mathrm{Categorical}(w)$
\State \hspace{\algorithmicindent}for each selected $\bz$: simulate $\bx\!\sim\!p(\bx\mid\bz)$, append $(\bz,\bx)$ to $\mathcal{D}_\mathrm{live}$, store $\log q'_\phi(\bz | \bx_0)\!\gets\!\log q_\phi(\bz_i\mid\bx_0)$
\State \hspace{\algorithmicindent}\textbf{return} $\mathcal{D}_\mathrm{live}$
\State \textbf{end function}

\Statex
\State \textbf{function} \texttt{posterior\_inference}$\big(q_\phi,\,q_\psi,\,\bx_0,\,J\big)$ \Comment{final sampling after training}
\State \hspace{\algorithmicindent}draw $K$ candidates $\{\bz_k\}\!\sim\! q_\phi(\bz\mid\bx_0)$
\State \hspace{\algorithmicindent}compute $\ell_k^{\mathrm{aux}}=\log q_\psi(\bz_k)$
\State \hspace{\algorithmicindent}$\text{mask}_k\!=\!100$ if $\bz_k$ out of prior, else $0$
\State \hspace{\algorithmicindent}define $\log w_k=-\ell_k^{\mathrm{aux}}-\text{mask}_k$
\State \hspace{\algorithmicindent}normalise and resample $J$ points to obtain weighted posterior samples
\State \hspace{\algorithmicindent}\textbf{return} $\{\bz\}_{1:J}$
\State \textbf{end function}
\end{algorithmic}
\end{algorithm*}

\begin{algorithm*}
\caption{\textbf{Dynamic Simulation-Based Inference-B} (DS-B)}
\begin{algorithmic}[1]
\State \textbf{Input:}
\State \hspace{\algorithmicindent} (Implicit) prior distribution $p(\bz)$
\State \hspace{\algorithmicindent} (Implicit) likelihood model/simulator $p(\bx \mid \bz)$
\State \hspace{\algorithmicindent} Neural density estimator $q_\phi(\bz \mid \bx)$
\State \hspace{\algorithmicindent} Number of simulations $N$
\State \hspace{\algorithmicindent} Observation $\bxobs$
\State \hspace{\algorithmicindent} Proposal update function $\tilde{p}(\bz) \leftarrow q_\phi(\bz \mid \bxobs)$
\State \hspace{\algorithmicindent} Dataset update function $\mathcal{D}_{\mathrm{live}} \leftarrow \texttt{data\_update}[\mathcal{D}_{\mathrm{live}}, \bxobs, \tilde{p}(\bz)]$
\State \hspace{\algorithmicindent} Convergence criterion $\texttt{converged}[\mathcal{D}_{\mathrm{live}}, \bxobs, q_\phi(\bz \mid \bx), \tilde{p}(\bz)]$
\State \hspace{\algorithmicindent} Log-proposal quantile $\alpha_B$
\State \textbf{Output:} Approximate posterior $q_\phi(\bz \mid \bx)$
\State
\State \textbf{Initialisation:}
\State Initialise proposal distribution $\tilde{p}(\bz) = p(\bz)$
\State Initialise training dataset $\mathcal{D}_{\mathrm{live}} \gets \emptyset$
\For{$i = 1$ to $N$}
    \State Sample $\bz^{(i)}$ from the prior $p(\bz)$
    \State Generate synthetic data $\bx^{(i)}$ from the model $p(\bx \mid \bz)$
    \State Add the data $(\bx^{(i)}, \bz^{(i)})$ and weights $w_i = 1$ to the training dataset $\mathcal{D}_{\mathrm{live}}$
\EndFor
\State Initialise neural network $q_\phi(\bz \mid \bx)$ with parameters $\phi$
\State
\State \textbf{Main Algorithm:}
\While{not $\texttt{converged}[\mathcal{D}_{\mathrm{live}}, \bxobs, q_\phi(\bz \mid \bx), \tilde{p}(\bz)]$} \textbf{(in parallel)}
    \State \textbf{[Parallel Process 1] Training Loop:}
    \For{$i = 1$ to $N_\mathrm{batches}$}
        \State Sample mini-batch $\mathcal{B}_i$ from $\mathcal{D}_{\mathrm{live}}$
        \State Compute the training loss: $\mathcal{L}_\mathrm{train}[\phi, \mathcal{B}_i] = -(1/\mid\mathcal{B}_i\mid) \, \sum_{j = 1}^{\mid\mathcal{B}_i\mid}{w_j \log q_\phi(\bz_j \mid \bx_j)}$
        \State Update network parameters $\phi$ using gradient descent on $\mathcal{L}_\mathrm{train}[\phi, \mathcal{B}_i]$
    \EndFor
    \State Update proposal distribution $\tilde{p}(\bz)  \leftarrow q_\phi(\bz \mid \bxobs) $
    \State
    \State \textbf{[Parallel Process 2] Update Step:}
    \State Update dataset $\mathcal{D}_{\mathrm{live}} \leftarrow \texttt{data\_update}[\mathcal{D}_{\mathrm{live}}, \bxobs, \tilde{p}(\bz)]$
    \State \textbf{function} \texttt{data\_update}[$\mathcal{D}_{\mathrm{live}}, \bxobs, \tilde{p}(\bz)$]
    \State \hspace{\algorithmicindent} Estimate critical threshold $c_B$ with current $\tilde{p}(\bz)$ given $\alpha_B$
    \State \hspace{\algorithmicindent} Delete all points $\bz_i$ in $\mathcal{D}_{\mathrm{live}}$ with $\tilde{p}(\bz_i) < c_B$
    \State \hspace{\algorithmicindent} Resample each $\bz_i \sim \tilde{p}(\bz)$ and $\bx_i \sim p(\bx_i \mid \bz_i)$
    \State \hspace{\algorithmicindent} Store $(\bz_i, \bx_i, w_i = p(\bz_i) / \tilde{p}(\bz_i))$ in $\mathcal{D}_{\mathrm{live}}$
    \State \textbf{end function}
\EndWhile
\State
\State \textbf{Return} Trained neural network $q_\phi(\bz \mid \bx)$
\end{algorithmic} \label{alg:OS-B}
\end{algorithm*}

\section{Implementation Details of DS-A}
\label{appendix:os_a_details}

\noindent All experiments in Section~\ref{sec:benchmarking} used Neural Spline Flows (NSFs)~\cite{Durkan2019nsf} as implemented in \texttt{sbi}~\cite{TejeroCantero2020sbi} to parameterise a conditional density $q_{\phi}(\mathbf{z}\mid\mathbf{x})$. 
Below, we detail the exact preprocessing and normalisation pipeline we implemented.

\begin{itemize}[leftmargin=*]
    \item \textbf{Hypercube preprocessing via inverse CDF mapping.} Let $\mathbf{z}^{\text{phys}}\!\in\!\mathbb{R}^D$ denote the parameters in their \emph{physical} units. We first map each component to a compact, common domain via the prior’s inverse CDF.  For the bimodal benchmark we employ per–dimension \texttt{uniform} priors: $z^{\text{phys}}_k\!\sim\!\mathcal{U}([\ell_k,h_k])$. With $F_k(z)=\cfrac{z-\ell_k}{h_k-\ell_k}$, the unit–interval representation is
    \begin{equation}
    u_k \;=\; F_k\!\big(z^{\text{phys}}_k\big) \;\in\; [0,1],
    \end{equation}
    which we rescale to a hypercube
    \begin{equation}
    \hat z_k \;=\; 4\,u_k - 2 \;\in\; [-2,2].
    \end{equation}
    Collecting coordinates gives $\mathbf{u}\!=\!(u_1,\ldots,u_D)$ and $\hat{\mathbf{z}}\!=\!(\hat z_1,\ldots,\hat z_D)$. This stage ensures that: (a) all rescaled parameters occupy a comparable dynamic range, (b) the prior’s per–dimension geometry is preserved, and (c) posterior mass is concentrated in a compact box where the downstream flow is most expressive and numerically stable to train.
    \item \textbf{Online normalisation for parameters and features.} Before passing parameters to the flow, we apply an online standardisation to the \emph{hypercube} variables $\hat{\mathbf{z}}$. Let $\boldsymbol{\mu}\in\mathbb{R}^D$ and $\boldsymbol{\sigma}\in\mathbb{R}^D_{>0}$ be the running mean and (diagonal) standard deviation tracked during training. We define the \emph{normalised parameter}:
    \begin{equation}
    \mathbf{z}^{\text{norm}} \;=\; \cfrac{\hat{\mathbf{z}} - \boldsymbol{\mu}}{\boldsymbol{\sigma}}
    \qquad(\text{elementwise}).
    \end{equation}
    The conditional flow takes $\mathbf{z}^{\text{norm}}$ as its random input and a summary $\mathbf{s}$ of the data (see below), \emph{i.e.},\ it models $\log p\!\left(\mathbf{z}^{\text{norm}}\mid\mathbf{s}\right)$. Since we ultimately require the likelihood in the \emph{pre-normalised} coordinates $\hat{\mathbf{z}}$, the negative log-likelihood must include the Jacobian of the affine change of variables:
    \begin{equation}
    \begin{aligned}
    &\Bigl|\det\!\Bigl(\cfrac{\partial \mathbf{z}^{\mathrm{norm}}}{\partial \hat{\mathbf{z}}}\Bigr)\Bigr|
    = \prod_{k=1}^D \cfrac{1}{\sigma_k} ,\\
    &\log\Bigl|\det\!\Bigl(\cfrac{\partial \hat{\mathbf{z}}}{\partial \mathbf{z}^{\mathrm{norm}}}\Bigr)\Bigr|
    = \sum_{k=1}^D \log \sigma_k \;\equiv\; \log(\mathrm{volume}) .
    \end{aligned}
    \end{equation}
    Hence, the loss we minimise includes a sum of terms of the form,
    \begin{equation}
        -\log p\!\left(\hat{\mathbf{z}}\mid \mathbf{s}\right)
    \;=\;
    -\log p\!\left(\mathbf{z}^{\text{norm}}\mid \mathbf{s}\right)
    \;+\; \log(\mathrm{volume}).
    \end{equation}
    
    This correction is \emph{probabilistically exact} and ensures that the normalisation improves optimisation conditioning without altering the density in the original (pre–normalised) coordinates.
    
    \medskip
    We process the observed features $\mathbf{x}^{\text{phys}}\!\in\!\mathbb{R}^{d_x}$ analogously, using a separate \texttt{LazyOnlineNorm} module (with its own running statistics) to obtain $\mathbf{x}^{\text{norm}}$. 
    A lightweight linear embedding then forms the conditional summary
    \begin{equation}
        \mathbf{s} \;=\; \mathbf{W}\,\mathbf{x}^{\text{norm}} + \mathbf{b},
    \quad \mathbf{W}\in\mathbb{R}^{2D\times d_x},\ \mathbf{b}\in\mathbb{R}^{2D}.
    \end{equation}
    Because $\mathbf{s}$ serves only as \emph{conditioning} input (not as a variable whose density is evaluated), no Jacobian term is required for this step.
    \item \textbf{NSF tail behaviour and why we further contract inputs.} In \texttt{sbi}/\texttt{nflows}, the rational–quadratic spline is applied on a bounded interval $[-B,B]$ (the \texttt{tail\_bound}, default $B{=}3$). Inside this region ($| v | \!\le\!B$), it is highly nonlinear but switches to \emph{linear tails} outside to guarantee a well-conditioned bijection. During training, transient covariate shifts can push a non-negligible fraction of activations outside $[-B,B]$, reducing effective expressivity there. To keep most mass within the nonlinear window, we apply an additional internal contraction before the flow:
    \begin{equation}
        \mathbf{v} \;=\; \alpha\,\mathbf{z}^{\text{norm}}, 
        \qquad \alpha = 0.2.
    \end{equation}
    This is fully accounted for by a change-of-variables term in the loss and thus does not bias the learned density; it simply increases the fraction of inputs that benefit from the spline’s nonlinear capacity. 
    \item \textbf{Optimisation.} We used the AdamW~\cite{loshchilov2019decoupledweightdecayregularization} optimiser with batch size $128$ and initial learning rate $10^{-2}$. The learning rate was adapted with a \texttt{ReduceLROnPlateau} module (with mode=\texttt{min}) monitored on the validation loss. We set the \emph{patience} to $10$ epochs, meaning the learning rate is reduced by a factor $\eta$ (here $\eta=\texttt{lr\_decay\_factor}$) \emph{only after} $10$ consecutive epochs without an improvement in the monitored metric.
    In the proposal-tempering step, we follow
\[
    \tilde p(\bz)\propto p(\bxobs\mid\bz)^{\gamma}p(\bz),\
    \tilde p(\bz)=\Big(\tfrac{q_\phi(\bz\mid\bxobs)}{p(\bz)}\Big)^{\frac{\gamma}{1+\gamma}}p(\bz),
\]
and (consistent with the empirical insensitivity of the posterior to moderate tempering within $\gamma\!\in\![0.1,1]$) we fix the tempering factor to $\gamma=0.5$ in our experiments.
    Training proceeded for $\sim 45$ minutes and used $\sim 4\times 10^{4}$ total simulations.\footnote{Experiments were run on a single NVIDIA GeForce RTX 3080~Ti (12\,GB), driver~570.172.08, CUDA~12.8.}
    \item \textbf{Summary:} (i) The inverse-CDF hypercube mapping respects the prior and removes gross scale disparities; (ii) online standardisation stabilises optimisation under streaming data and yields the exact loss correction via the Jacobian volume term;  (iii) the small contraction in the penultimate bullet point keeps activations inside the spline’s nonlinear window $| v | \!\le\!B$, improving effective expressivity while preserving likelihood correctness via a change of variables. Together, these choices yield stable, data-efficient online learning without sacrificing posterior coverage.
\end{itemize}

\begin{table}
    \centering
    \renewcommand{\arraystretch}{1.2}
    \begin{tabular}{c c c c c c c}
        \hline
        Component & Parameter & True value & Initial prior  \\
        \hline
        \parbox[t]{1mm}{\multirow{6}{*}{\rotatebox[origin=c]{90}{SPLE}}}
        & $x_\mathrm{lens}\, ['']$ & -0.05 & $\mathcal{U}(-0.2, 0.2)$  \\
        & $y_\mathrm{lens}\, ['']$ & 0.10 & $\mathcal{U}(-0.2, 0.2)$ \\
        & $\phi_\mathrm{lens} \, [\mathrm{rad}]$ & 1.0 & $\mathcal{U}(0.5, 1.5)$ \\
        & $q_\mathrm{lens}$ & 0.75 & $\mathcal{U}(0.5, 1.0)$ \\
        & $r_\mathrm{ein}\, ['']$ & 1.5 & $\mathcal{U}(1, 2)$ \\
        \hline
        \parbox[t]{1mm}{\multirow{2}{*}{\rotatebox[origin=c]{90}{Shear}}}
        & $\gamma_1$ & 0.005 & $\mathcal{U}(-0.5, 0.5)$  \\
        & $\gamma_2$ & -0.100 & $\mathcal{U}(-0.5, 0.5)$ \\
        \hline
        \parbox[t]{1mm}{\multirow{7}{*}{\rotatebox[origin=c]{90}{Source light}}}
        & $x_\mathrm{src}\, ['']$ & 0.0 & $\mathcal{U}(-0.2, 0.2)$ \\
        & $y_\mathrm{src}\, ['']$ & 0.0 & $\mathcal{U}(-0.2, 0.2)$ \\
        & $\phi_\mathrm{src} \, [\mathrm{rad}]$ & 0.75 & $\mathcal{U}(0.5, 1.25)$ \\
        & $q_\mathrm{src}$ & 0.5 & $\mathcal{U}(0.2, 0.8)$ \\
        & $n_\mathrm{src}$ & 2.3 & $\mathcal{U}(1.5, 3.0)$ \\
        & $r_e\, ['']$ & 2.0 & $\mathcal{U}(0.5, 3.0)$ \\
        & $I_e$ & 0.6 & $\mathcal{U}(0.1, 2.0)$ \\
        \hline
    \end{tabular}
    \caption{True parameter values and priors used in the strong gravitational lensing application.}
    \label{tab:lensing-params}
\end{table}

\clearpage

\end{document}